\documentclass[twocolumn]{aastex631}
\usepackage{amsmath}
\usepackage{graphicx}
\usepackage{placeins}
\usepackage{float}
\usepackage{txfonts}
\usepackage{hyperref}

\usepackage{lipsum}
\newcommand{\sw}[1]{\texttt{#1}}
\newcommand\blfootnote[1]{%
  \begingroup
  \renewcommand\thefootnote{}\footnote{#1}%
  \addtocounter{footnote}{-1}%
  \endgroup
}

\clearpage

\newcommand{\Co}{$\rm ^{56}Co$}
\newcommand{\Fe}{$\rm ^{56}Fe$}

\begin{document}

\title{Characterizing the Ordinary Broad-lined Type Ic SN 2023pel from the Energetic GRB 230812B }

\correspondingauthor{Gokul P. Srinivasaragavan}\email{gsriniv2@umd.edu}
\author[0000-0002-6428-2700]{Gokul P. Srinivasaragavan}
\affiliation{Department of Astronomy, University of Maryland, College Park, MD 20742, USA}
\affiliation{Joint Space-Science Institute, University of Maryland, College Park, MD 20742, USA}
 \affiliation{Astrophysics Science Division, NASA Goddard Space Flight Center, 8800 Greenbelt Rd, Greenbelt, MD 20771, USA}

 \author[0000-0002-7942-8477]{Vishwajeet Swain}
\affiliation{Department of Physics, Indian Institute of Technology Bombay, Powai, 400 076, India}

\author[0000-0002-9700-0036]{Brendan O'Connor}
    \affiliation{McWilliams Center for Cosmology, Department of Physics, Carnegie Mellon University, Pittsburgh, PA 15213, USA}
    
\author[0000-0003-3768-7515]{Shreya Anand}
\affiliation{Division of Physics, Mathematics and Astronomy, California Institute of Technology, Pasadena, CA 91125, USA}

\author[0000-0002-2184-6430]{Tomás Ahumada}
\affiliation{Division of Physics, Mathematics and Astronomy, California Institute of Technology, Pasadena, CA 91125, USA}

\author[0000-0001-8472-1996]{Daniel Perley}
\affiliation{Astrophysics Research Institute, Liverpool John Moores University, Liverpool Science Park, 146 Brownlow Hill, Liverpool L3 5RF, UK}

\author[0000-0003-2434-0387]{Robert Stein}
 \affiliation{Division of Physics, Mathematics and Astronomy, California Institute of Technology, Pasadena, CA 91125, USA}

\author[0000-0003-1546-6615]{Jesper Sollerman}
\affiliation{Department of Astronomy, The Oskar Klein Center, Stockholm University, AlbaNova, 10691 Stockholm, Sweden}

\author[0000-0002-4223-103X]{Christoffer Fremling}
\affiliation{Caltech Optical Observatories, California Institute of Technology, Pasadena, CA 91125, USA}
\affiliation{Division of Physics, Mathematics and Astronomy, California Institute of Technology, Pasadena, CA 91125, USA}

\author[0000-0003-1673-970X]{S. Bradley Cenko}
\affiliation{Astrophysics Science Division, NASA Goddard Space Flight Center, 8800 Greenbelt Rd, Greenbelt, MD 20771, USA}
\affiliation{Joint Space-Science Institute, University of Maryland, College Park, MD 20742, USA}

\author[0000-0002-7686-3334]{S.~Antier}
\affiliation{Observatoire de la C\^ote d'Azur, Universit\'e C\^ote d'Azur, Boulevard de l'Observatoire, 06304 Nice, France}

\author[0000-0003-1585-8205]{Nidhal Guessoum}
\affiliation{Physics Department, American University of Sharjah, Sharjah, UAE}

\author[0009-0009-2434-432X]{Thomas Hussenot-Desenonges}\affiliation{IJCLab, Univ Paris-Saclay, CNRS/IN2P3, Orsay, France}

\author{Patrice Hello}
\affiliation{IJCLab, Univ Paris-Saclay, CNRS/IN2P3, Orsay, France}

\author[0000-0001-8058-9684]{Stephen Lesage}
\affiliation{Center for Space Plasma and Aeronomic Research, University of Alabama in Huntsville, Huntsville, AL 35899, USA}
\affiliation{University of Alabama in Huntsville, 320 Sparkman Drive, Huntsville, AL 35899, USA}

\author[0000-0002-5698-8703]{Erica Hammerstein}
\affiliation{Department of Astronomy, University of Maryland, College Park, MD 20742, USA}

\author[0000-0002-2666-728X]{M. Coleman Miller}
\affiliation{Department of Astronomy, University of Maryland, College Park, MD 20742, USA}
\affiliation{Joint Space-Science Institute, University of Maryland, College Park, MD 20742, USA}

\affiliation{Astrophysics Science Division, NASA Goddard Space Flight Center, 8800 Greenbelt Rd, Greenbelt, MD 20771, USA}
\affiliation{Center for Research and Exploration in Space Science and Technology, NASA/GSFC, Greenbelt, MD 20771, USA}

\author[0000-0002-8977-1498]{Igor Andreoni$^\dag$}
\blfootnote{$^\dag$Gehrels Fellow}
\affiliation{Joint Space-Science Institute, University of Maryland, College Park, MD 20742, USA}
\affiliation{Department of Astronomy, University of Maryland, College Park, MD 20742, USA}
 \affiliation{Astrophysics Science Division, NASA Goddard Space Flight Center, 8800 Greenbelt Rd, Greenbelt, MD 20771, USA}
 
\author[0000-0002-6112-7609]{Varun Bhalerao}
\affiliation{Department of Physics, Indian Institute of Technology Bombay, Powai, 400 076, India}

\author[0000-0002-7777-216X]{Joshua S. Bloom}
\affiliation{Department of Astronomy, University of California,
Berkeley, CA 94720-3411, USA}
\affiliation{Lawrence Berkeley National Laboratory, 1 Cyclotron Road,
MS 50B-4206, Berkeley, CA 94720, USA}

\author[0000-0002-7708-3831]{Anirban Dutta}
\affiliation{Indian Institute of Astrophysics, 2nd Block 100 Feet Rd, Koramangala Bangalore, 560 034, India}

\author[0000-0002-3653-5598]{Avishay Gal-Yam}
    \affiliation{Department of Particle Physics and Astrophysics, Weizmann Institute of Science, 76100 Rehovot, Israel}

\author[0000-0002-0129-806X]{K-Ryan Hinds}
\affiliation{Astrophysics Research Institute, Liverpool John Moores University, Liverpool Science Park, 146 Brownlow Hill, Liverpool L3 5RF, UK}

\author[0000-0002-3850-6651]{Amruta Jaodand}
\affiliation{Division of Physics, Mathematics and Astronomy, California Institute of Technology, Pasadena, CA 91125, USA}

\author[0000-0002-5619-4938]{Mansi Kasliwal}
\affiliation{Division of Physics, Mathematics and Astronomy, California Institute of Technology, Pasadena, CA 91125, USA}

\author[0000-0003-0871-4641]{Harsh Kumar}
\affiliation{Harvard College Observatory, Harvard University, 60 Garden St. Cambridge 02158 MA}
\affiliation{Department of Physics, Indian Institute of Technology Bombay, Powai, 400 076, India}

\author[0000-0002-2715-8460]{Alexander S. Kutyrev}
\affiliation{Department of Astronomy, University of Maryland, College Park, MD 20742, USA}
\affiliation{Astrophysics Science Division, NASA Goddard Space Flight Center, 8800 Greenbelt Rd, Greenbelt, MD 20771, USA}

\author[0000-0003-2132-3610]{Fabio Ragosta}
\affiliation{INAF, Osservatorio Astronomico di Roma, via Frascati 33, I-00078
Monte Porzio Catone (RM), Italy}

\author[0000-0002-7252-5485]{Vikram Ravi}
\affiliation{Division of Physics, Mathematics and Astronomy, California Institute of Technology, Pasadena, CA 91125, USA}

\author[0000-0002-4477-3625]{Kritti Sharma}
\affiliation{Division of Physics, Mathematics and Astronomy, California Institute of Technology, Pasadena, CA 91125, USA}

\author[0000-0002-0525-0872]{Rishabh Singh-Teja}
\affiliation{Indian Institute of Astrophysics, 2nd Block 100 Feet Rd, Koramangala Bangalore, 560 034, India}

\author[0000-0002-2898-6532]{Sheng Yang}
\affiliation{Henan Academy of Sciences, Zhengzhou 450046, Henan, China}

\author[0000-0003-3533-7183]{G.C. Anupama}
\affiliation{Indian Institute of Astrophysics, 2nd Block 100 Feet Rd, Koramangala Bangalore, 560 034, India}

\author[0000-0001-8018-5348]{Eric C. Bellm}
\affiliation{DIRAC Institute, Department of Astronomy, University of Washington, 3910 15th Avenue NE, Seattle, WA 98195, USA}

\author[0000-0002-8262-2924]{Michael W. Coughlin}
\affiliation{School of Physics and Astronomy, University of Minnesota,
Minneapolis, Minnesota 55455, USA}

\author[0000-0003-2242-0244]{Ashish~A.~Mahabal}
\affiliation{Division of Physics, Mathematics and Astronomy, California Institute of Technology, Pasadena, CA 91125, USA}
\affiliation{Center for Data Driven Discovery, California Institute of Technology, Pasadena, CA 91125, USA}

\author[0000-0002-8532-9395]{Frank J. Masci}
\affiliation{IPAC, California Institute of Technology, 1200 E. California
             Blvd, Pasadena, CA 91125, USA}

\author[0009-0002-7897-6110	]{Utkarsh Pathak}
\affiliation{Department of Physics, Indian Institute of Technology Bombay, Powai, 400 076, India}

\author[0000-0003-1227-3738]{Josiah Purdum}
\affiliation{Caltech Optical Observatories, California Institute of Technology, Pasadena, CA  91125}

\author[0000-0002-7150-9061]{Oliver J. Roberts}
\affiliation{Science and Technology Institute, Universities Space and Research Association, 320 Sparkman Drive, Huntsville, AL 35805, USA.}

\author[0000-0001-7062-9726]{Roger Smith}
\affiliation{Caltech Optical Observatories, California Institute of Technology, Pasadena, CA  91125}

\author[0000-0002-9998-6732]{Avery Wold}
\affiliation{IPAC, California Institute of Technology, 1200 E. California
             Blvd, Pasadena, CA 91125, USA}

\begin{abstract}
We report observations of the optical counterpart of the long gamma-ray burst (LGRB) GRB 230812B, and its associated supernova (SN) SN 2023pel. The proximity ($z = 0.36$) and high energy ($E_{\gamma, \rm{iso}} \sim 10^{53} \, \rm{erg}$) make it an important event to study as a probe of the connection between massive star core-collapse and relativistic jet formation. With a phenomenological power-law model for the optical afterglow, we find a late-time flattening consistent with the presence of an associated SN. SN 2023pel has an absolute peak $r$-band magnitude of $M_r = -19.46 \pm 0.18$ mag (about as bright as SN 1998bw) and evolves on quicker timescales. Using a radioactive heating model, we derive a nickel mass powering the SN of $M_{\rm{Ni}} = 0.38 \pm 0.01 \, \rm{M_\odot}$, and a peak bolometric luminosity of $L_{\rm{bol}} \sim 1.3 \times 10^{43}\,  \rm{erg \, s^{-1}}$. We confirm SN 2023pel's classification as a broad-lined Type Ic SN with a spectrum taken 15.5 days after its peak in $r$ band, and derive a photospheric expansion velocity of $v_{\rm{ph}} = 11,300 \pm 1,600 \, \rm{km\,s^{-1}}$ at that phase. Extrapolating this velocity to the time of maximum light, we derive the ejecta mass  $M_{\rm{ej}} = 1.0 \pm 0.6 \, \rm{M_\odot}$ and kinetic energy $E_{\rm{KE}}  = 1.3^{+3.3}_{-1.2} \times10^{51} \, \rm{erg}$. We find that GRB 230812B/SN 2023pel has SN properties that are mostly consistent with the overall GRB-SN population. The lack of correlations found in the GRB-SN population between SN brightness and $E_{\gamma, \rm{iso}}$ for their associated GRBs, across a broad range of 7 orders of magnitude, provides further evidence that the central engine powering the relativistic ejecta is not coupled to the SN powering mechanism in GRB-SN systems.
\end{abstract}

\section{Introduction}
\label{sec:intro}
A clear link has been established over the past two decades between long-duration gamma-ray bursts (LGRBs; $T_{90} > $ 2 s) and core-collapse supernovae (CCSNe) from an observational basis \citep{Woosley2006}. Photometrically, a characteristic SN ``bump" arises in the afterglow light curve (LC) within 10 to 20 days, as the afterglow fades. Over 40 LGRBs with this characteristic bump have been discovered (see, e.g., \citealt{Hjorth2013, cano2017,Melandri2019, Hu2021, Kumar2022, Rossi2022, Blanchard2023, Srinivasaragavan2023}), and are known as GRB-SNe. Spectroscopic observations of these SNe have revealed that almost all (SN 2011kl associated with GRB 111209A was a super-luminous SN; \citealt{Kann2011, Greiner2015}) are broad-lined Type Ic SNe (Type Ic-BL; \citealt{Woosley2006}); they lack hydrogen and helium lines in their optical spectra, and have broad lines corresponding to ejecta velocities higher than those seen in normal Type Ic explosions. Before this work, 28 GRB-SNe have been spectroscopically confirmed (see, e.g.,  \citealt{cano2017, Cano2017b, Wang2018b, Melandri2019, Hu2021, Kumar2022, Rossi2022, Blanchard2023}).


However, there remain a number of open questions surrounding the GRB-SN connection, and recent discoveries have shown that our understanding of the connection may not be as complete as once thought. A SN is not always detected for nearby LGRBs \citep{GalYam2006, Fynbo2006, DelleValle2006, Tanga2018}, and the physical link between LGRBs and their associated SN is also not clear. Studies of the brightest GRB of all time, GRB 221009A \citep{Lesage+2023,Frederiks+2023,LHAASO+2023,Malesani+2023,Burns2023,Williams+2023,Laskar+2023,Kann+2023,OConnor+2023}, have shown that its associated SN has a peak luminosity consistent with those of the rest of the GRB-SN population \citep{Blanchard2023, Srinivasaragavan2023, Fulton2023, Shrestha+2023, Kann+2023, Levan2023}, despite the GRB being more luminous by orders of magnitude. On the other hand, SN 1998bw associated with GRB 0980425 was a very nearby ($z = 0.0085$; \citealt{Iwamoto98, 98bwpaper, patat2001}) and relatively luminous SN, with a derived nickel mass ($M_{\rm{Ni}}$) powering the SN as high as 0.9 $\rm{M_\odot}$ \citep{Sollerman2000}. Its associated GRB 980425 was a low-luminosity GRB, with an isotropic equivalent energy of $E_{\gamma, \rm{iso}} \sim 10^{48} \, \rm{erg}$ \citep{Galama1998}, which is three to four orders of magnitude fainter than what is seen for cosmological GRBs. Numerous observational studies have also been done on GRB-SNe whose associated GRBs have energies in between GRB 221009A and GRB 0908425 (see, e.g.,  \citealt{Malesani2004, Mazzali2006, Matheson2003, Starling2011, Schulze2014}), and they paint a scattered picture regarding the relationship between GRB energetics and SN properties. 

The origin and classification of LGRBs based solely on their $T_{90}$ has also come into question, as GRB 211211A ($T_{90} \sim 34.3 \, \rm{s}$; \citealt{31210GCN}) and GRB 230307A ($T_{90} \sim 35 \, \rm{s}$; \citealt{GCN33411}) may have had associated kilonova emission, pointing towards a compact object origin \citep{Rastinejad2022, Troja2022, Yang2022, Levan2023b, Yang2023, Gillanders2023}. Studies of LGRBs' surrounding interstellar medium also show evidence for LGRBs that do not arise from the collapse of massive stars, but rather compact object mergers as well \citep{Lesniewska2022}. Therefore, it is important to follow up nearby, bright LGRBs and their associated SNe across the electromagnetic spectrum in order to shed light on some of these questions surrounding the GRB-SN connection. Here we report on the characterization of one such GRB-SN.

GRB 230812B was discovered by the \textit{Fermi} Gamma-Ray Burst Monitor (GBM; \citealt{Meegan+2009}) at 18:58:12 UTC on 12 August 2023, which we establish hereafter as $\rm{T_0}$ \citep{GCN34387}. The burst has a $T_{90} = 2.95 \pm 1.02 $ s, and a  fluence of $2.69\pm 0.01 \, \times 10^{-4} \, \rm{erg \, cm^{-2}}$ in the 10--1000 keV band. The afterglow was subsequently detected as an X-ray point source by the \textit{Swift} X-Ray Telescope (XRT; \citealt{Burrows2005}) at $\rm{T_0} + 0.297$ days \citep{GCN34400}.  Its brightness prompted follow-up across the electromagnetic spectrum \citep{OpticalGCN1, OpticalGCN2, OpticalGCN3, GCNZTF}, and the optical counterpart was discovered by \citet{GCN34395} and \citet{GCNZTF}. Spectroscopic observations of the optical afterglow led to a redshift measurement of $z=0.36$ \citep{GCNredshift}. At this redshift using a flat $\Lambda$CDM cosmology with $\Omega_{\rm m}=0.286$ and H$_{0} = 69.6$~km~s$^{-1}$~Mpc$^{-1}$ to convert redshifts to distances, the burst has an isotropic equivalent energy release of $E_{\gamma, \rm{iso}} \sim 1.1 \times 10^{53} \, \rm{erg}$. Using the $T_{90}$ reported, the burst has an isotropic equivalent average gamma-ray luminosity of $L_{\gamma, \rm{iso}} \sim 8.8 \times 10^{52} \, \rm{erg\, s^{-1}}$. Comparing these values to the LGRB population with observationally confirmed SN, GRB 230812B possesses the fifth highest $E_{\gamma, \rm{iso}}$ and second highest $L_{\gamma, \rm{iso}}$. This makes GRB 230812B a rare example of an energetic LGRB nearby enough to search for an associated Type Ic-BL SN.


In this $\textit{Letter}$, we present optical observations of the afterglow of GRB 230812B that display a clear late-time flattening consistent with an associated SN (SN 2023pel; \citealt{SNGCN}), and spectroscopic observations confirming SN 2023pel as a Type Ic-BL SN.
In \S \ref{Observations} we report our observations of the optical counterpart; in \S\ref{agsn} we analyze the optical counterpart and find its associated SN 2023pel; in \S \ref{SNanalysis} we analyze SN 2023pel and characterize its key properties; and in \S \ref{Conclusion} we summarize our conclusions. \citet{Hussennot2023} also report an analysis of this event, and where relevant we compare our results with theirs.

\section{Observations} 
\label{Observations}

\begin{figure*}
    \centering
    \includegraphics[width = \linewidth]{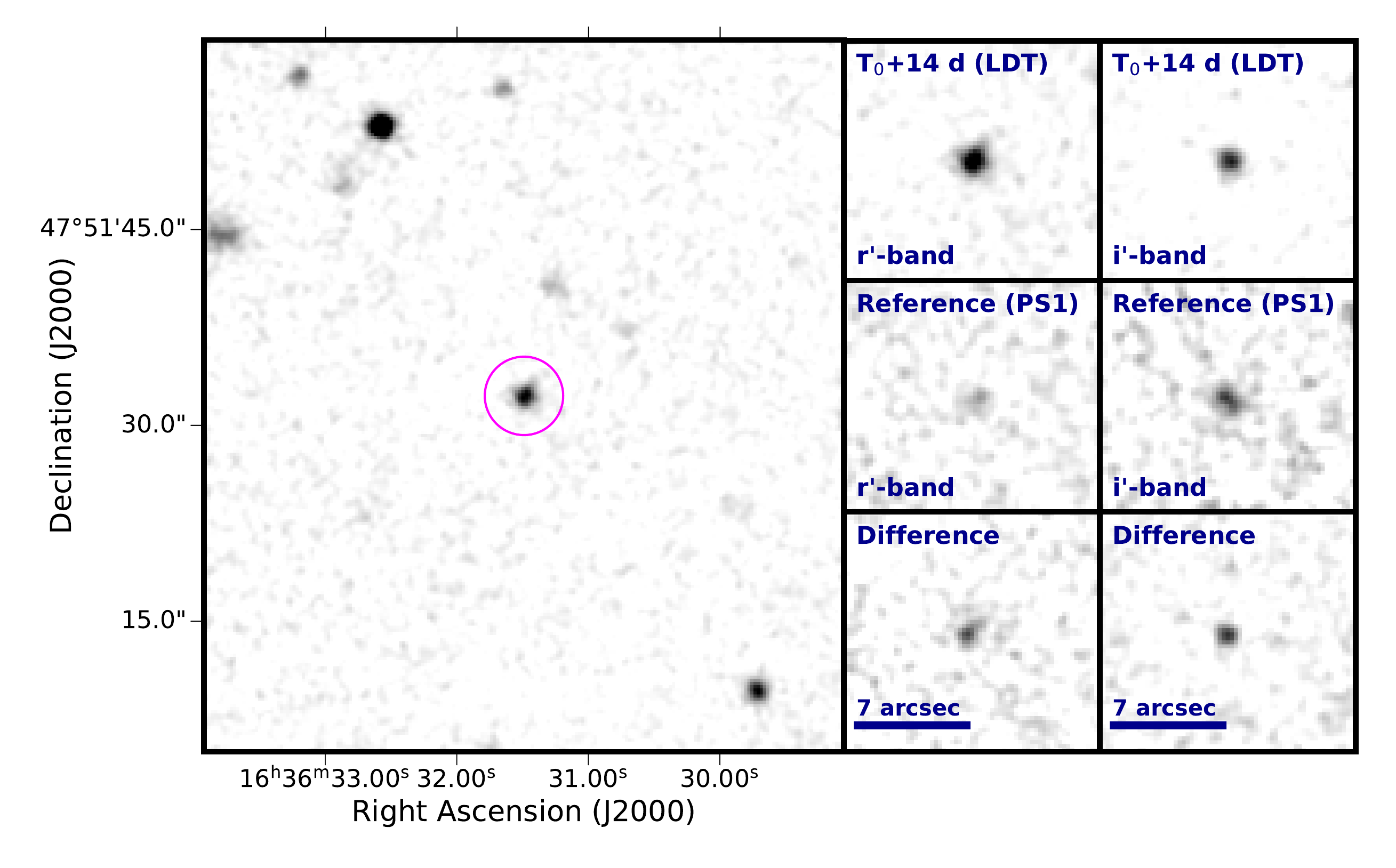}
    \caption{LDT images of the optical afterglow of GRB 230812B/SN 2023pel. The left, large panel shows the wider field of view of GRB 230812B, in $r$ band 14 days after $\rm{T_0}$. The right panels show the observations taken of the GRB that include the host galaxy contribution, the reference images from PS1 used for image subtraction, and the resulting subtracted images used for photometry, in both $r$ and $i$ bands. The images have been smoothed for display purposes.}
    \label{LDTobs}
\end{figure*}

\subsection{\textit{Swift}}

Observations of the afterglow with the \textit{Neil Gehrels Swift Observatory} \citep{Gehrels_2004} XRT began at $\rm{T_0}+25$ ks \citep{GCN34400}, localizing the afterglow of the GRB. The data were obtained in Photon Counting (PC) mode. We retrieved the time-averaged XRT PC mode spectrum from the \textit{Swift}-XRT GRB Lightcurve Repository\footnote{\url{https://www.swift.ac.uk/xrt_curves/}} \citep{Evans2009}. The spectrum contains 5.0 ks of data obtained between $\rm{T_0}+25$ ks to $\rm{T_0}+38$ ks, with a midtime of $\rm{T_0}+8.8$ hr. The data are grouped to a minimum of 1 count per bin. We also retrieve the X-ray LC, which contains data obtained between $\rm{T_0} + 25$ ks to $\rm{T_0} + 1421$ ks.

We likewise retrieved the data obtained by the Ultra-Violet/Optical Telescope \citep[UVOT;][]{Roming2005}
from the \textit{Swift} Data Archive\footnote{\url{https://heasarc.gsfc.nasa.gov/cgi-bin/W3Browse/swift.pl}}. All observations were obtained using the broadband \textit{white} filter (hereafter, \textit{wh}). 
We focused on the initial \textit{wh}-filter data (ObsID: 00021619001) with a total exposure of 4,881 s. The data were combined using the \texttt{uvotimsum} task within \texttt{HEASoft v6.29c} \citep{HEASOFT}. Due to an issue with the \textit{Swift} gyroscope affecting attitude control \citep{GCN34633} the image point spread function (PSF) is elongated. 
We therefore utilized a lenticular source region with position angle (PA) of $302^\circ$ 
to match the shape of the PSF and selected two nearby, source-free circular regions of 15\arcsec radius as background. We used the \texttt{uvotsource} and \texttt{uvot2pha} task to measure the photometry and retrieve the spectral files. At this epoch, we measure a source brightness of $wh=19.29\pm0.06$ AB mag.
We obtain similar results (consistent within $1\sigma$ errors) using \texttt{uvotdetect}, which automatically selects the source region. Therefore, we conclude that our treatment of the elongated PSF is reasonable.

\subsection{ZTF}
The Zwicky Transient Facility (ZTF; \citealt{Bellm2019,Dekany2020,Graham2019, Masci2019})  is a public-private survey that images the entire northern sky every two days in the $g$ and $r$ bands. The transient detection with ZTF relies on image subtraction, using templates of predefined fields. Any high significance difference ($>5\sigma$) generates an \textit{alert} containing information about the transient \citep{ZTfalert}. We query the alert stream via Kowalski \citep{kowalski} through an alert filtering scheme on Fritz previously described in SGRB and gravitational-wave searches \citep{Ahumada+2021,ahumada2022,kasliwal2020kilonova}. In a nutshell, we select sources that are spatially and temporally consistent with the GBM localization, are far from PS1 stars (based on \citealt{TaMi2018}) or bright sources, are real based on the real-bogus score \citep{kowalski}, have a positive residual, and that have at least two detections separated by a minimum of 15 minutes. 

We ingested the GBM localization map \citep{GbmDataTools} into Fritz \citep{VanderWalt2019}, the ZTF instance of SkyPortal \citep{skyportal2019, Coughlin2023skyportal}, an interactive tool design to plan and schedule ToO observations for ZTF. The observing plan was generated using \texttt{gwemopt} \citep{CoAn2019}, by taking the healpix \textit{Fermi}-GBM localization map, slicing the skymap into predefined tiles of the size and shape of the ZTF field of view, determining which fields have the highest enclosed probability, and optimizing observations based on airmass and visibility windows. For this purpose, we used a modified version of the ‘greedy’ algorithm \citep{CoTo2018,almualla2020scheduling}, only allowing for the use of the ZTF primary grid. The final schedule consisted of 300 s exposures in  the $r$ and $g$ bands, starting 8.6 hours after the GBM detection. The observing plan for the first night covered 420 square degrees beginning at 2023-08-13 03:34:57, using 3 epochs of 9 fields, totaling 2.25 hours. This corresponds to 78\% of the probability enclosed in the Earth-occultation corrected GRB localization map. The exposures reached median depths of 21.9 mag in both $g$ and $r$ band \citep{GCNZTF}. The first ZTF detection of the afterglow ZTF23aaxeacr happened 8.61 hours after the burst, during the first ZTF exposure of the field, as the transient was sitting in the ZTF field that covered the highest probability. 

For GRB 230812B, ZTF detected 22154 sources in the GBM error region in difference imaging, though only 55 sources passed our filtering criteria. These sources were cross-matched against the \textit{Wide-field Infrared Survey Explorer} (\textit{WISE}; \citealt{2013wise.rept....1C}), milliquas \citep{milliquas}, and the  Minor Planet Center, to ensure these were not active galactic nuclei \citep{WISEcut} or solar system objects. Finally, we queried the IPAC-ZTF forced photometry service using a month-long baseline to reject young SNe. The majority of the candidates were ruled out by one (or multiple) of the criteria previously described. The afterglow of GRB 230812B, ZTF23aaxeacr/\hbox{SN\,2023pel} [$\alpha$ (J2000)= 16$^{\mathrm{h}}$36$^{\mathrm{m}}$31.48$^{\mathrm{s}}$,
$\delta$ (J2000) = +47$^{\circ}$51$\arcmin$32$\farcs$26; \citealt{GCNZTF}], was found during the first night of observations due to the fast evolution of 2 mag day$^{-1}$ showed in the r-band ZTF data. No other candidate showed a photometric evolution consistent with an afterglow, and the source was reported to TNS (AT 2023pel; \citealt{GCNZTF}) once its fast evolution was confirmed. Our TNS report came only 4 minutes after the refined \textit{Swift} XRT localization, and our source was consistent (within 6 arcsec) with the center of the XRT region \citep{GCN.34394}. 

\subsection{SEDM}

Once the afterglow was identified, we used the Rainbow Camera on the Spectral Energy Distribution Machine (SEDM; \citealt{BlNe2018, Rigualt2019}) mounted on the Palomar 60-inch telescope to acquire $u-$, $g-$, $r-$, and $i-$ band imaging in 300\,s exposures. The SEDM images started 11.02 hours after the burst. The SEDM images were processed with a \texttt{Python}-based pipeline version of the \texttt{Fpipe} \citep{FrSo2016}, which includes photometric calibrations and image subtraction using reference images from The Sloan Digital Sky Survey (SDSS; \citealt{SDSS}).





\subsection{LT}
The location of the GRB was observed with IO:O, the optical imager on the 2m robotic Liverpool Telescope \citep{Steele2004} at the Observatorio del Roque de los Muchachos.  Observations were taken on three separate nights:  2023-08-14 ($griz$ filters), 2023-08-16 ($gri$), and 2023-09-04 ($r$).

Reduced images were downloaded from the LT archive and processed with custom image-subtraction and analysis software (K. Hinds and K. Taggart et al., in prep.)  Image stacking and alignment is performed using {\sc SWarp} \citep{SWarp} where required.  Image subtraction is performed using a pre-explosion reference image in the appropriate filter from the Panoramic Survey Telescope and Rapid Response System 1 (PS1; \citealt{Flewelling2016}). The photometry is measured using PSF fitting methodology relative to PS1 standards and is based on techniques in \cite{FrSo2016}. 

\subsection{LDT}
We also observed GRB 230812B's optical counterpart in $r$ and $i$ with the 4.3m Large Monolithic Imager (LMI) on the Lowell Discovery Telescope (LDT) for 5 epochs between UT 2023-08-26 and 2023-09-09. We reduced the images using a custom \texttt{Python}-based image analysis pipeline \citep{Toy2016}, that performs data reduction, astrometry, registration, source extraction and PSF photometry using \sw{SourceExtractor} \citep{SourceExtractor}. \sw{SourceExtractor} was calibrated using point-sources from PS1 DR1 catalog. We then performed image subtraction using the PS1 templates and the \sw{ZOGY} algorithm-based \texttt{Python} pipeline \citep{PyZogy, Kumar2022} to remove the host contribution.
Figure \ref{LDTobs} shows both the wider field of view of GRB 230812B's position on the sky, as well as its flux in both filters at $T_0 + 14$ days.

\subsection{GIT}
\label{GIT}
We used the 0.7m GROWTH-India Telescope (GIT) \citep{2022AJ....164...90K} located at the Indian Astronomical Observatory (IAO), Hanle-Ladakh, to acquire data of GRB 230812B's optical counterpart. The counterpart was observed in Sloan $g^\prime$, $r^\prime$ and $i^\prime$ bands starting 20~h after $\rm{T_0}$. We continued observations for up to seven days by acquiring multiple 300 s exposures. Data were downloaded and processed in real time by the GIT data reduction pipeline \citep{2022MNRAS.516.4517K}. We used individual exposures of 300 s for photometry in the early stages when the afterglow was bright. Later, we stacked images with \sw{SWarp} \citep{2010ascl.soft10068B} to increase the S$/$N ratio of detections.\\

All images were pre-processed by subtracting bias images, flat-fielding, and removing cosmic-rays via Astro-SCRAPPY \citep{2019ascl.soft07032M} package. Astrometry was performed on the resulting images using the offline \sw{solve-field} \citep{2010AJ....139.1782L} astrometry engine. Subsequently, refined astrometry was conducted using the \sw{SCAMP} \citep{2006ASPC..351..112B} package to facilitate image stacking with \sw{SWarp} \citep{2010ascl.soft10068B}. Sources were detected using \sw{SourceExtractor} \citep{1996AAS..117..393B} and were crossed-matched against the PS1 DR1 catalog \citep{2016arXiv161205560C} through Vizier to obtain the zero-point in images. Using the \sw{ZOGY} algorithm-based \texttt{Python} pipeline, we performed image subtraction on all the images using the PS1 templates. Finally, the pipeline performed point spread function (PSF) fit photometry on subtracted images to obtain magnitudes.

\subsection{HCT}
The 2-meter Himalayan Chandra Telescope (HCT) situated at the Indian Astronomical Observatory (IAO) in Hanle-Ladakh was used to observe the optical counterpart of GRB 230812B. This counterpart was observed in Sloan $r^\prime$ and $i^\prime$ bands, beginning 3 days after $\rm{T_0}$. During the period of UT 2023-08-15 to 2023-08-24, we conducted four observations, capturing multiple exposures lasting between 20 to 40 minutes each. Standard image reduction techniques were applied, including bias subtraction and flat-fielding, as well as cosmic-ray removal using the Astro-SCRAPPY  package. Astrometry was conducted on the resulting images using the offline \sw{solve-field} astrometry engine. We then use the same methods as \S \ref{GIT} to extract sources, perform image subtraction, and perform PSF photometry to get magnitudes of the GRB counterpart.


\subsection{Keck}
\label{DEIMOS}
We obtained a spectrum of GRB 230812B/SN 2023pel on UT 2023-09-09 05:31:47 using the Deep Imaging Multi-Object Spectrograph (DEIMOS; \citealt{DEIMOS}) mounted on the 10-m Keck II telescope. Our configuration used a 600ZD grating, a central wavelength of 7500 $\AA$ and the OG 550 filter to maximize redder wavelength coverage. The observation consisted of 3 $\times$ 1800s exposures and one 900s exposure totalling 1.75 hr. The data were calibrated and reduced using \sw{Pypeit} \citep{PypeIt}.


\section{Afterglow Analysis}
\label{agsn}

\subsection{Characterization of X-ray and Optical Afterglow}
\label{aganalysis}


We start with determining some of the basic afterglow properties, and note that an in-depth, multi-wavelength study of the afterglow is forthcoming in Pathak et al. (in preparation). We find that the power-law temporal decay indices of the X-ray (\textit{Swift} XRT) and optical data (using $g$, $r$ and $i$-bands prior to when the SN emission affects the LC, $< T_0 + 4$ days), are consistent, with $\alpha_X = 1.31^{+0.07}_{-0.06}$ ($\chi^2_{\rm{reduced}} = 0.96$),  and $\alpha_O = 1.31 \pm 0.02$ ($\chi^2_{\rm{reduced}} = 2.14$). We then calculate the spectral index in the optical, using SEDM observations in the $u$, $g$, $r$, and $i$ bands from SEDM at $\rm{T_0} + 10.8$ hours. We derive $\beta_O = 0.74 \pm 0.02$ ($\chi^2_{\rm{reduced}} = 0.76$) after correcting for line-of-sight extinction through the Galactic plane ($A_V = 0.06$ mag; \citealt{Schlafly2011}).
These values are consistent with the results of \citet{Hussennot2023}, who find $\alpha_O = 1.35 \pm 0.02$ and $\beta_O = 0.74 \pm 0.01$. 

\begin{figure}
    \centering
    \includegraphics[width = \linewidth]{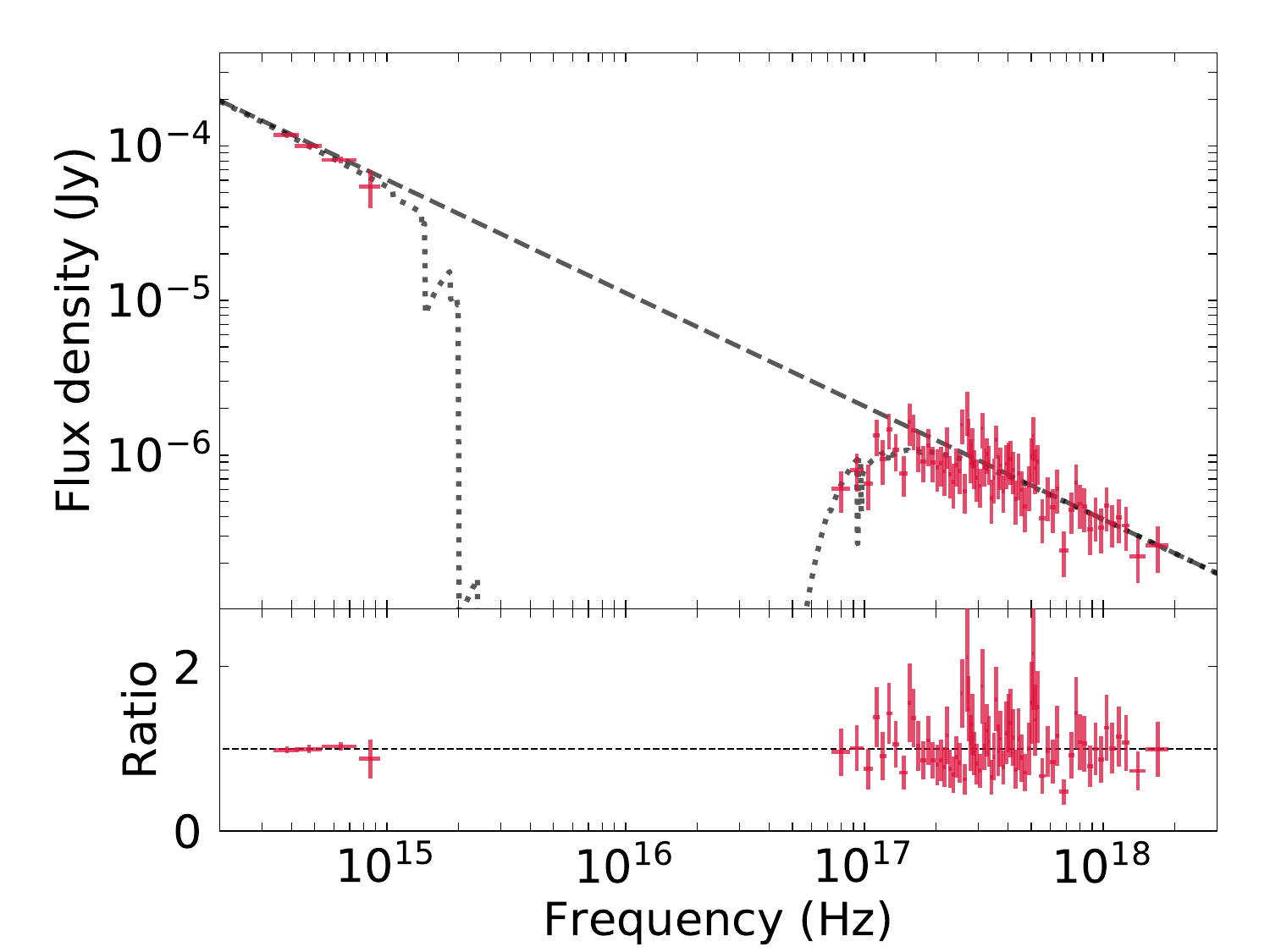}
    \caption{Broadband modeling of the GRB afterglow from optical (ZTF and SEDM; $ugri$) to X-ray (\textit{Swift}) wavelengths at a midtime of $\rm{T_0} + 8.8$ hours. The model is a simple absorbed power-law shown by the gray dotted line and the unabsorbed model is shown as the gray dashed line, both corrected for Galactic extinction. The bottom panel shows the ratio between the data and the absorbed power-law model with the black dashed line showing a ratio of unity.}
    \label{broadbandmodeling}
\end{figure}

We then perform an X-ray analysis through \texttt{XSPEC v12.12.0} using the initial PC mode XRT spectrum. We modelled the spectrum assuming an absorbed power-law model \texttt{tbabs*ztbabs*pow}, which is a Tuebingen-Boulder interstellar medium (ISM) absorption model that calculates the cross-section for X-ray absorption by the ISM \citep{Absorptionmodel}. We fix the Galactic hydrogen column density to $N_{H,\textrm{MW}}=2.0\times10^{20}$ cm$^{-2}$ \citep{Willingale2013} and a redshift of $z=0.36$. We fit the data by minimizing Cash statistics \citep{Cash1979}. We obtain a best-fit (C-stat = 259 for 332 dof) X-ray photon index of $\Gamma_X=1.765\pm0.085$ and intrinsic hydrogen column density $N_{H,z}=(1.2\pm0.4)\times10^{21}$ cm$^{-2}$. This corresponds to an X-ray spectral index of $\beta_X = \Gamma_X - 1 = 0.765 \pm 0.085$. The consistency between the optical and X-ray spectral indices suggests that the optical and X-ray data lie on the same spectral segment (see Figure \ref{broadbandmodeling}). 

Therefore, we include the early optical data ($u$, $g$, $r$, and $i$ filters) obtained by SEDM, each shifted to a midtime of 8.8 hr using the best-fit temporal power-law, to constrain the possibility of dust intrinsic to the GRB environment, through modeling the broadband spectral energy distribution (SED) again in \texttt{XSPEC v12.12.0}. We fix the Milky Way dust reddening to  $E(B-V)_\textrm{MW}=0.02$ mag \citep{Schlafly2011}. The broadband SED was fit using the model \texttt{tbabs*ztbabs*redden*zdust*pow}, which again uses the Tuebingen-Boulder ISM absorption model to account for absorption and extinction both in the Milky Way and the host galaxy. We applied a Milky Way extinction law with $R_V=3.1$ \citep{Cardelli1989}, and derive a photon index of $\Gamma_{XO}=1.73\pm0.02$, $N_{H,z}=(1.1\pm0.2)\times10^{21}$ cm$^{-2}$, and $E(B-V)_z<0.07$ mag (3$\sigma$) 
for a $\chi^2=363$ for 336 dof. This is consistent with the results of \citet{Hussennot2023}, who find $E(B-V)_z<0.03$ mag. We show the broadband modeling described, along with the X-ray and optical data at a midtime of $\rm{T_0} + 8.8$ hours in Figure \ref{broadbandmodeling}. Given the low upper limit derived, we ignore the host galaxy extinction for rest of our analysis.

The spectral index we derive ($\beta_{\rm{OX}} = 0.73 \pm 0.02$) does not match with a locally fast cooling environment ($\nu_m<\nu_c<\nu$, where $\nu_m$ is the injection frequency of the electrons and $\nu_c$ is the cooling frequency), as the slope of the electron energy distribution $p$ would be abnormally low ($1.46\pm0.04$), according to the standard closure relations \citep{Sari1998,Granot2002}. Instead, a more reasonable value of $p=2.46\pm0.04$ is obtained for $\nu_m<\nu<\nu_c$, assuming an adiabatic jet and constant density ISM. This points towards a slow-cooling regime for the synchrotron afterglow in the optical to X-ray bands (see Pathak et al., in preparation, for more details).

\begin{figure*}
    \centering
    \includegraphics[width = \linewidth]{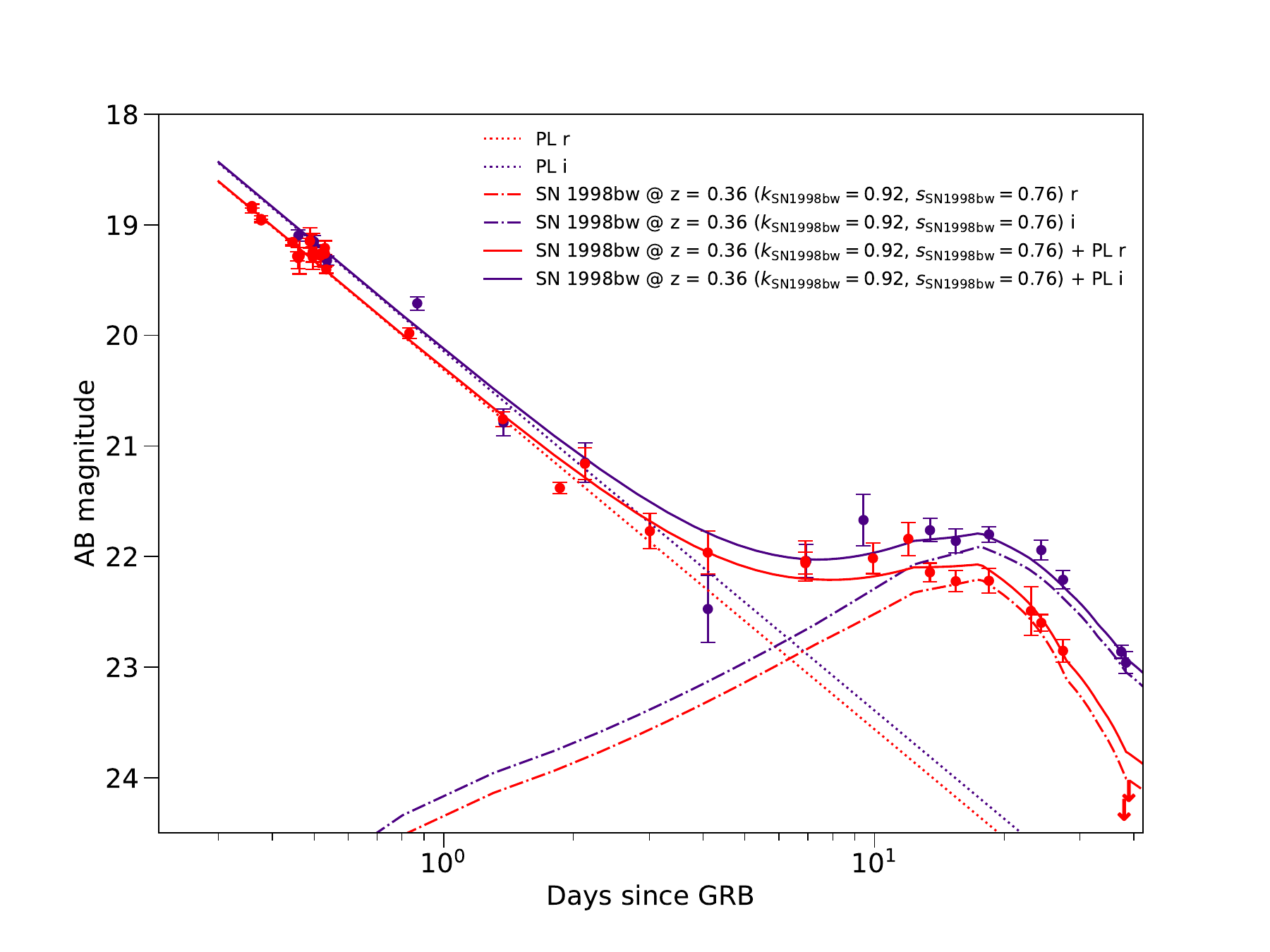}
    \caption{Observed $r$ and $i$ band photometry of GRB 230812B from the optical instruments listed in \S \ref{Observations}, with all magnitudes already host-galaxy subtracted. The photometry values are reported in Table \ref{Table1}, and all times are in the observer frame. We also show the best-fit power law decay to the optical data derived in \S \ref{aganalysis}, corresponding to a power-law decay index of $\alpha = 1.31$. The LCs for a SN 1998bw-like source in $r$ and $i$ bands, redshifted to $z=0.36$, reddened according to the line-of-sight Galactic extinction of GRB 230812B, with a flux-stretching factor of $k=0.92$ and time-stretching factor of $s=0.76$ (derived in \S \ref{SNbump}) applied, are also shown. }
    \label{agphotometry}
\end{figure*}

\subsection{SN Bump}
\label{SNbump}
A clear late-time flattening is seen as soon as $\rm{T_0} + 7$ days in Figure \ref{agphotometry}, in both the $r$ and $i$ bands, which we interpret as due to rising SN emission. \citet{Kann2007} and \citet{Oates2009} have shown that if the central engine is still active, a re-brightening of the optical afterglow may occur shortly after the prompt emission. \citet{Greiner2009} has also shown that optical flaring due to refreshed or reversed shocks can cause brightening episodes in the early-time optical afterglow LC. However, the X-ray LC of GRB 230812B (see Figure \ref{magnetarmodel}) shows no evidence of flaring at early times, and does not re-brighten when the optical afterglow begins to, which occurs a week after the prompt emission. Therefore, we determine that the late-time flattening cannot be explained by central engine activity. 

It has been shown that the LGRBs 211211A and 230307A have evidence of associated kilonova emission \citep{Rastinejad2022, Troja2022, Yang2022, Levan2023b}. In order to determine if the late-time flattening could be due to a kilonova, we transform the kilonova LC associated with GRB 130603B \citep{Berger2013} to the redshift of GRB 230812B ($z=0.36$), and apply the line-of-sight Galactic extinction. We find that the kilonova peaks around 26 mag in $H$ band, and therefore, would be even fainter in the optical bands. This is too faint to cause any significant late-time flattening in the optical afterglow LC, which leads us to determine that the late-time flattening cannot be explained by an associated kilonova.

Therefore, the most likely explanation for the late-time flattening is an associated SN. In order to compare the associated SN to SN 1998bw (the prototypical SN associated with a GRB used in modeling studies, due to the proximity of the SN at $z = 0.0085$), we use the \sw{PyMultiNest} Bayesian modeling package \citep{Feroz09, Buchner+2014} to find the best-fit flux-stretching factor ($k_{\rm{SN1998bw}}$) and time-stretching factor ($s_{\rm{SN1998bw}}$) of the SN with respect to SN 1998bw. The full model that we use is: 
\begin{equation}
 f_\nu({t_{\rm{obs}}}) = k_{\rm{SN1998bw}} (f_\nu^{\rm{SN1998bw}}(t_{\rm{obs}}/s_{\rm{SN\, 1998bw}})) + a\\_{\rm{AG}} (t_{\rm{obs}})^{-\alpha},
 \end{equation} 
where $ f_\nu^{\rm{SN1998bw}}(t_{\rm{obs}})$ is the flux seen of SN 1998bw at $z = 0.36$ at a time in the observer frame, $t_{\rm{obs}}$ is the time in the observer frame, $\alpha$ is the power-law decay index, and $a_{\mathrm{AG}}$ is the flux constant of proportionality. We derive $ f_\nu^{\rm{SN1998bw}}$ by (de)reddening and K-correcting the nugent-hyper model \citep{Levan2005} in \sw{SNCosmo} \citep{SNCosmo} to match the relevant properties of GRB 230812B. We perform the fitting concurrently in the $r$ and $i$ bands, fixing $\alpha = 1.31$ and $a_{\rm{AG}}$ (derived in \S \ref{aganalysis}). Therefore, the free parameters in our fitting procedure corresponding are $k_{\rm{SN1998bw}}$ and $s_{\rm{SN1998bw}}$. Because a possible correlation may exist between these two parameters \citep{Cano2014}, we create priors for both of the parameters drawn from a bivariate normal distribution fit to the $k_{\rm{SN1998bw}}$ and $s_{\rm{SN1998bw}}$ values derived for GRB-SN in literature \citep{cano2017}. Furthermore, in order to account for systematic uncertainties due to combining data from multiple telescopes (S-corrections; \citealt{Stritzinger2002}), we numerically optimize the likelihood function assuming that the reported errors actually underestimate the true uncertainty. We do this through using the same method as \citet{Srinivasaragavan2023}, by introducing an error-stretching factor $\beta$ in the fitting procedure to represent the S-correction. 

We find that the best-fit values for the flux-stretching and time-stretching factors are $k_{\rm{SN1998bw}} = 0.92$ and $s_{\rm{SN1998bw}} = 0.76$, with median $\pm \, 1\sigma$ values of $k_{\rm{SN1998bw}} = 0.93^{+0.04}_{-0.03}$ and $s_{\rm{SN1998bw}} = 0.76 \pm 0.02$. The reduced chi-squared statistic is $\chi^2_{\rm{reduced}} = 1.2$, indicative that the model is an adequate fit to the observed data. Our $k_{\rm{SN1998bw}}$ is consistent with the value found by \citet{Hussennot2023} of $k_{\rm{SN1998bw}} = 1.04 \pm 0.09$, and our $s_{\rm{SN1998bw}}$ is consistent at the 2$\sigma$ level with theirs of $s_{\rm{SN1998bw}} = 0.68 \pm 0.05$. Therefore, we find that the SN associated with GRB 230812B, SN 2023pel, is about as bright as SN 1998bw, but evolves on a quicker timescale. We will revisit this when we model the SN parameters in \S \ref{Arnett}. We note that the presence of a jet break in the optical LC is a possible source of systematic error for this analysis; however, the lack of a jet break in the X-ray LC and our well-sampled optical data set make this an unlikely possibility.

After subtracting the best-fit power law value from the brightest $r$-band photometry point seen in the late-time flattening and correcting for the Galactic extinction, we find that the observed peak absolute magnitude of SN 2023pel is $M_r = -19.46 \pm 0.18$ mag, which is consistent with the peak magnitude found in \citet{Hussennot2023}, $M_r = -19.41 \pm 0.10$. As expected from the flux-stretching factor we derived, this is consistent with the peak absolute magnitude of SN 1998bw, $M_R = -19.36 \pm 0.05$ mag \citep{Galama1998}, and brighter than what is seen for the the overall Type Ic-BL SN population ($M_r = -18.6 \pm 0.5$ mag; \citealt{Taddia2018}). We confirm that SN 2023pel is indeed a Type Ic-BL SN in \S \ref{spectra}.

\subsection{Spectrum Analysis}
\label{spectra}
\begin{figure*}
    \centering
     \includegraphics[width = 0.9\linewidth]{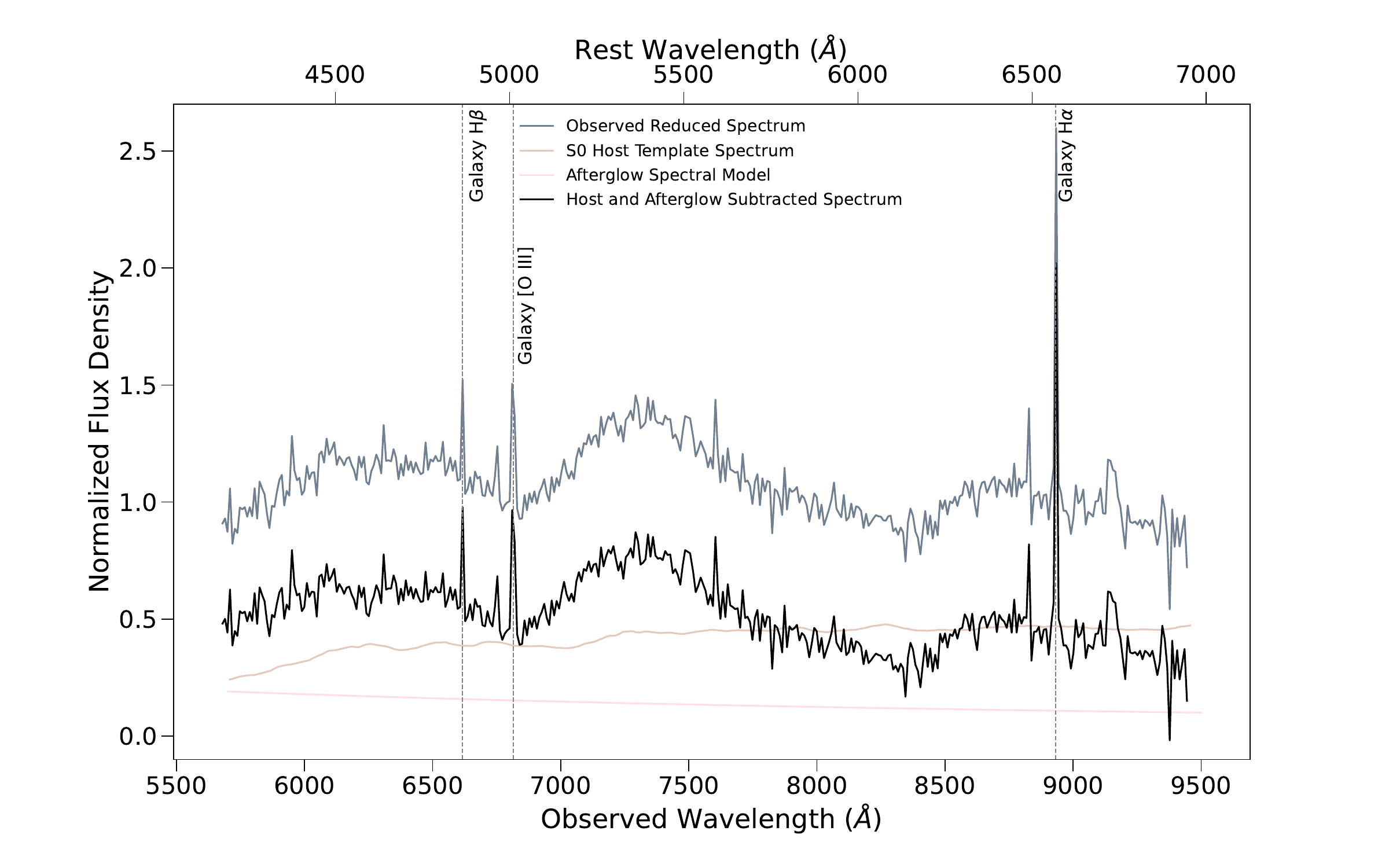}
    \includegraphics[width = 0.9\linewidth]{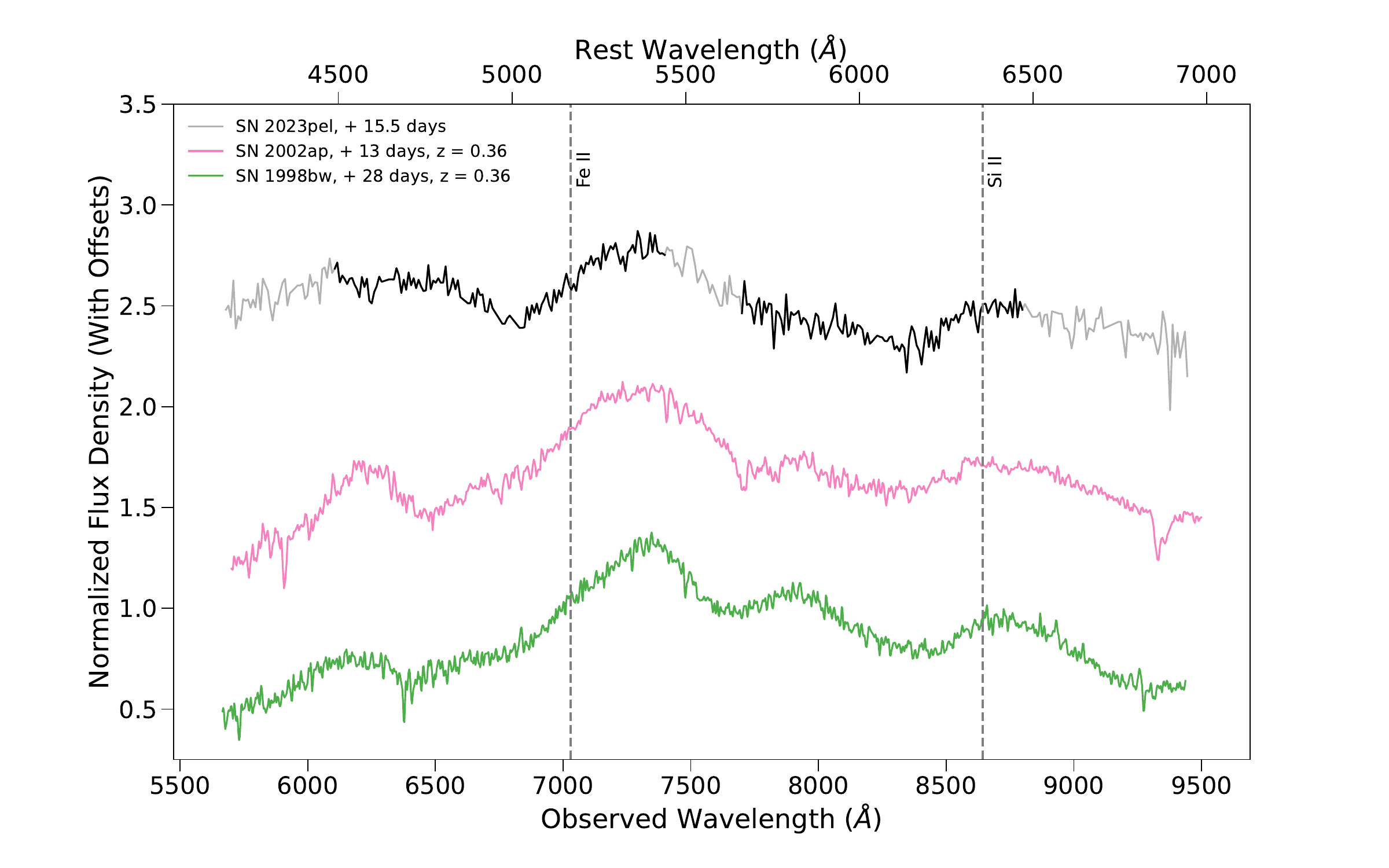}
    \caption{\textit{Top panel}: DEIMOS spectrum obtained of GRB 230812B/SN 2023pel, 15.5 days after the observed peak in $r$ band. We show the original reduced spectrum, the best-fit S0 host galaxy template spectrum from NGSF with its flux scaled to the time of the spectrum, the afterglow spectral model with its flux scaled to the time of the spectrum assuming $F_\nu \propto \nu^{-\beta_O}$ (where $\beta_O = 0.74$ is the optical spectral index), and the final, host and afterglow-subtracted spectrum of SN 2023pel, smoothed and normalized for display purposes. The spectrum shows the strong H$\beta$, H$\alpha$, and [O III] galaxy emission lines at $z=0.36$. \textit{Bottom panel}: We show the spectrum of SN 2023pel, after manually removing narrow galaxy emission lines. We also show the observed spectrum of SN 2002ap, 13 days after peak, and redshifted to $z=0.36$, and of SN 1998bw, 28 days after peak, also redshifted to $z=0.36$. All three spectra show broad Fe II and Si II features, characteristic of Type Ic-BL SNe, and we label the lines as well as show the broadened features in bold for SN 2023pel. }
    \label{SNSpectrum}
\end{figure*}

As described in \S \ref{DEIMOS}, we obtained a DEIMOS spectrum on UT 2023-09-09 05:31:47 of GRB 230812B/SN 2023pel, and we show the observed, reduced spectrum after correcting for telluric features in gray in the top panel of Figure \ref{SNSpectrum}. $\rm{H}\beta$, $\rm{H}\alpha$, and [O III] galaxy emission lines are clearly seen in the spectrum, and we use {\sc PySpecKit}'s {\sc splot} interactive fitting routine \citep{Pyspeckit1, Pyspeckit2} to fit these lines to a redshift. We find $z = 0.36112 \pm 0.00004$  (where the error is the root mean square error), providing an independent confirmation of the redshift of GRB 230812B/SN 2023pel. 

We then use the Next Generation Super Fitter (NGSF; \citealt{ngsf}) to model the SN and host together. For this, we allow NGSF to explore all the available SN and galaxy templates, while fixing the value for the redshift and limiting the SN phase to a window between 5 and 25 days after peak. The results show that the NGSF best fit is to an Elliptical galaxy with a SN-Ic BL ($\chi^2/\rm{dof} \sim 7.45$). However, elliptical hosts for LGRBs are extremely rare, as they usually originate from active star-forming galaxies. So far, GRB 050219A is the only example of a LGRB found in an elliptical \citep{Rossi2014} galaxy, and the presence of narrow emission lines at the redshift of the host leads us to conclude that the host is most likely not an elliptical galaxy. Therefore, we use the next-best fit with a SN-Ic BL, a S0 spiral galaxy ($\chi^2/\rm{dof} \sim 9.80$), to represent the host contribution. We note that NGSF does not include nebular emission in their galaxy templates, though we mask the galaxy lines before performing the fits. When re-doing the analysis using the Elliptical galaxy template, all of our derived results were consistent within error bars. The particular template used does not seem to significantly impact our analysis.

We scale the host contribution to the spectrum to the same percentage (44\%) of the host contribution to the $r$-band photometry of GRB 230812B's optical counterpart at the time closest to the spectrum ($T_0 + 27.394$ days), derived from comparing the LDT magnitudes before and after image subtraction. We show the template in brown in the top panel Figure \ref{SNSpectrum}, and subtract this template from the observed spectrum. We also subtract the afterglow's spectral model, corresponding to $F_\nu \propto \nu^{-\beta_O}$, where $\beta_O = 0.74$ is the optical spectral index derived in \S \ref{aganalysis}, from the observed spectrum. We scale the spectral model of the afterglow to the same percentage (14\%) of the afterglow's contribution to the observed $r$-band photometry at the time of the spectrum, through using the best-fit temporal power-law derived in \S \ref{aganalysis}. We show the spectral model in pink in the top panel of Figure \ref{SNSpectrum}. 

The final host and afterglow-subtracted, smoothed, and normalized spectrum of SN 2023pel is shown in black in the top panel of Figure \ref{SNSpectrum}. The phase of the spectrum corresponds to 15.5 days after the observed peak in $r$ band. We then use the IDL routine \sw{WOMBAT} to remove the galaxy emission lines, and show that spectrum in the bottom panel of Figure \ref{SNSpectrum}. The spectrum shows clear broad absorption features characteristic of Type Ic-BL SNe, confirming the classification from \citet{classificationGCN}. We indicate the Fe II and Si II features in the Figure. We then run the SN Identification code (SNID; \citealt{Blondin2007}) to determine the best match Ic-BL templates. We find that the template for SN 2002ap, 13 days after its peak, and SN 1998bw, 28 days after its peak, are good matches to the spectrum. As mentioned in \S \ref{SNbump}, we determined that SN 2023pel evolves on a quicker timescale than SN 1998bw. This agrees with the best-fit SN 1998bw template SNID found, as the matching template has a phase 13 days later than the observed phase of the spectrum of SN 2023pel. We show the observed spectra of SN 2002ap \citep{Mazzali2002} and SN 1998bw \citep{patat2001} at the time of these best-fit templates, in pink and green in the bottom panel of Figure \ref{SNSpectrum}.

\section{SN Analysis}
\label{SNanalysis}
After showing evidence that a SN is the source of the late-time flattening of the afterglow LC, we shift our focus to the analysis of the SN itself. In this section, we investigate whether the SN can be powered by a millisecond magnetar (\S \ref{magnetar}), or if the observed data are better described by the conventional  \citet{Arnett1982} radioactive decay model (\S \ref{Arnett}), and report progenitor and SN properties corresponding to the respective models. We then analyze a spectrum taken of GRB 230812B/SN 2023pel (\S \ref{spectra}), and end by comparing the properties of GRB 230812B/SN 2023pel to those of the rest of the GRB-SN population (\S \ref{GRBSN}). 

\subsection{Magnetar Model}
\label{magnetar}
Although the \citet{Arnett1982} radioactive heating model (see \S \ref{Arnett}) is the frequently adopted model used to describe Type Ic-BL LCs, theories have shown that a millisecond magnetar central engine can power LGRBs \citep{Usov1992}, and observational studies of multiple GRB-SNe have suggested evidence for a magnetar origin \citep{Greiner2005, Kenji2007, Cano2016, Lu2018, Zhang2022}. We investigate whether our observed X-ray and optical LCs can be explained within the context of this model, using the formulation from \citet{Cano2016}, before using the \citet{Arnett1982} model. The millisecond magnetar model consists of three phases: an afterglow component
whose emission is due to the GRB ejecta colliding with the surrounding medium; a component whose emission is due to the powering of the central engine; and the SN component whose emission is also due to the powering of the central engine. The X-ray LC is described just by the first two phases, as the SN emission is not significant in the X-ray bandpasses, while the optical LC is described by the addition of all three phases. In order for this model to be viable, the initial  magnetic field strength ($B_0$) and spin period ($P_0$) derived from the X-ray and optical LC analysis must be consistent with each other. We note that this model assumes that the magnetic field is dipolar in nature, and unevolving over time. Therefore, there are no considerations for multipolar and rapidly evolving magnetic fields within the context of this model.

The full derivation of the model can be found in \citet{Cano2016}, and here we provide a brief description of the components. The afterglow is modeled by two components -- an impulsive energy input term, and a continuous energy input term. The impulsive energy input term \citep{Zhang2001} is represented by a simple power law (SPL; \citealt{Rowlinson2013}): 

\begin{equation}
L_{\rm SPL} (t) = \Lambda t^{-\alpha}
\label{equ:SPL}
\end{equation}

\noindent where $\Lambda$ is a normalisation constant and $\alpha$ is the power-law decay index. We assume that $\alpha = \Gamma_{\gamma} + 1$, where $\Gamma_{\gamma}$ is the photon index of the prompt emission. This definition of $\alpha$ comes from the assumption that the decay slope is governed by the curvature effect \citep{Panaitescu2000, Piran2004}. 

The continuous energy input term is due to the magnetar central engine depositing Poynting flux into the ejecta, where the neutron star is assumed to have a mass of $1.4 \, \rm{M_{\odot}}$ and a radius of $10^6$ cm. This emission creates a characteristic plateau in the LC, and is represented as: 

\begin{equation}
L_{\rm AG}(t)=L_{0}\left(1+\frac{t}{t_{0}}\right)^{-2} 
\label{equ:magnetar_AG}
\end{equation}

\noindent where $L_{0}$ is the luminosity of the plateau emission, and $t_{0}$ is the duration of the plateau. 

Finally, the SN component is modeled by the central engine depositing its energy directly into the SN after the initial jet spreads. The analytical prescription has been derived by many works \citep{ostriker1971, kasen2010, Barkov2011, Chatzopoulos2011}, and the equation \citep{Arnett1980, Arnett1982, Valenti2008, Chatzopoulos2009, Chatzopoulos2011} is represented as:

\begin{equation}
 L_{\rm SN}(t) = \frac{E_{\rm p}}{t_{\rm p}}~{\rm exp}\left(\frac{-x^2}{2}\right)\int_0^x~\frac{z~{\rm exp}\left(\frac{z^2}{2}\right)}{(1+yz)^2}\, \mathrm{d}z 
 \label{equ:mag_SN}
\end{equation}

\noindent where $E_{\rm p}$ is the initial energy of the magnetar in units of erg, $t_{\rm p}$ is the characteristic spin-down time of the magnetar in units of days, $x = t / t_{\rm diff}$, $y = t_{\rm diff} / t_{\rm p}$, and $t_{\rm diff}$ is the diffusion timescale of the SN in units of days. As mentioned earlier, all of this is considering an $l=2$ dipole. We can rewrite $E_{\rm p}$ and $t_{\rm p}$ in terms of $L_0$ and $t_0$ through 

\begin{equation}
E_{\rm p} = \frac{2.00}{2.05}~L_{0}t_{0}
\label{equ:E_LT_link}
\end{equation}

and

\begin{equation}
t_{\rm p} = 2~t_{0}  
\label{equ:tp_T_link}
\end{equation}
The final model we fit for the X-ray LC is 

\begin{equation}
 L_{\rm total}^{\rm{X-ray}} (t) = L_{\rm AG} + L_{\rm SPL}  
 \label{equ:mag_combined}
\end{equation}
and the optical LC, which we fit to the $r$ band, is 

\begin{equation}
 L_{\rm total}^{r} (t) = L_{\rm AG} + \Phi L_{\rm SN} + L_{\rm SPL}  
 \label{equ:mag_combined_norm_SN}
\end{equation}
\noindent where $\Phi$ is an additional free-parameter used to normalize the model to the optical data. Even if the properties of the magnetar derived from the X-ray and optical LCs match, if $\Phi > 1$, an additional source of energy is necessary to power the SN in addition to the magnetar central engine.

Finally, after fitting the observed LCs with the models presented, the $L_0$ and $t_0$ derived can be used to find $B_0$ and $P_0$ through

\begin{equation}
\frac{B_0}{10^{15}~\text{G}} = \sqrt{\frac{4.2}{L_{0,49}t^{2}_{0,3}}} 
\label{equ:mag_AG_B}
\end{equation}

and

\begin{equation}
\frac{P_0}{1~\text{ms}} = \sqrt{\frac{2.05}{L_{0,49}t_{0,3}}} 
\label{equ:mag_AG_P}
\end{equation}
\noindent where $L_{0,49} = L_{0} / 10^{49}$~erg~s$^{-1}$ and $t_{0,3} = t_{0} / 10^{3}$~s.

\begin{figure}
    \centering
    \includegraphics[width= \linewidth]{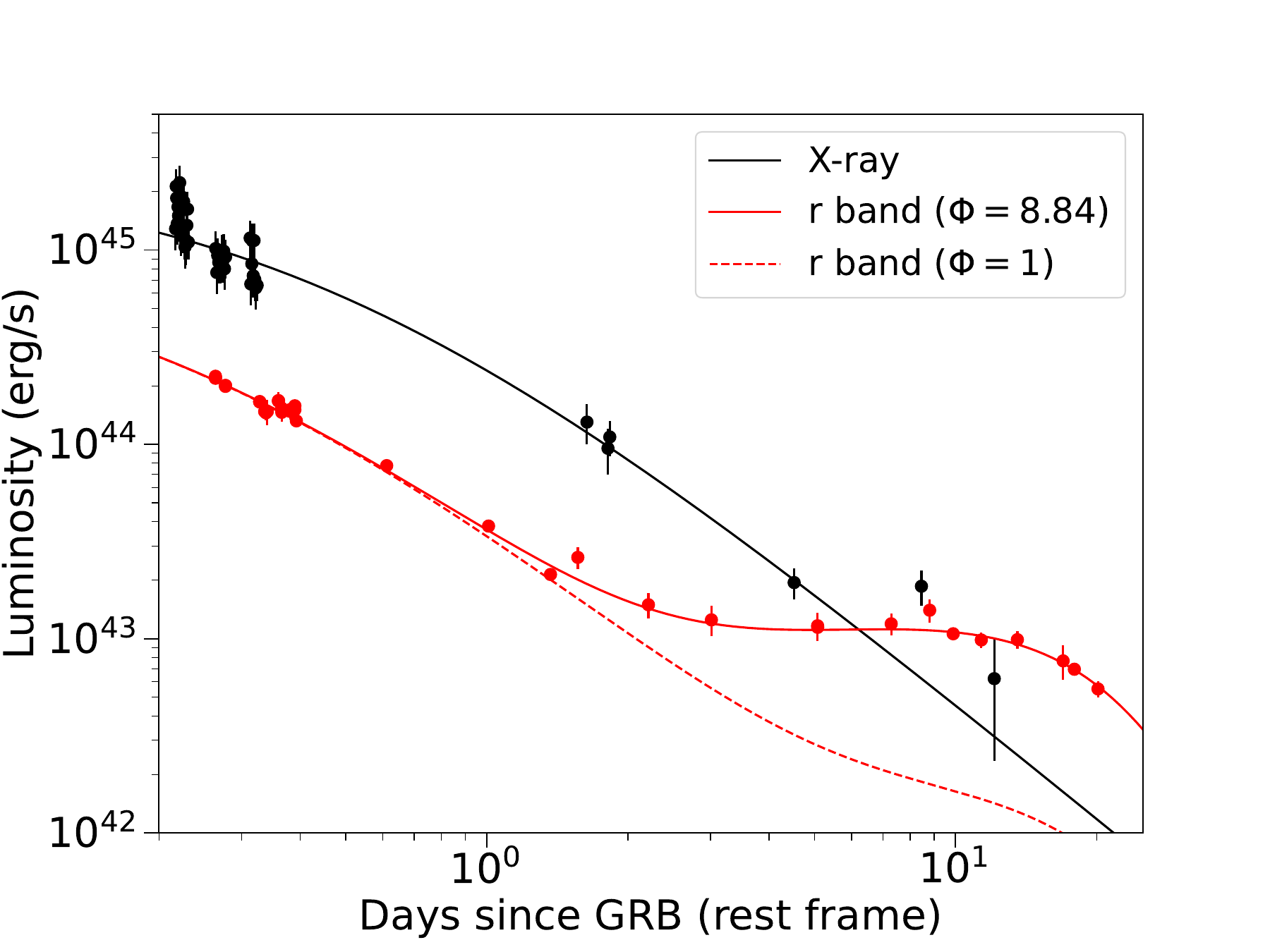}
    \caption{The magnetar model from \citet{Cano2016} fit to both the \textit{Swift} XRT X-ray LC and $r$-band LC of GRB 230812B/SN 2023pel independently. The properties of the magnetar derived are not consistent when comparing the X-ray and optical analysis, where we find that $B_0 = 3.37^{+0.91}_{-0.63} \times 10^{15} \, \rm{G}$ and $P_0 = 14.44^{+2.29}_{-1.93} $ ms from the X-ray fitting, and $B_0 = 1.07^{+0.13}_{-0.12} \times 10^{16} \, \rm{G}$ and $P_0 = 32.41^{+2.67}_{-2.48} $ ms from the optical fitting. We find that an additional flux-stretching factor of $\Phi = 8.84^{+0.62}_{-0.63}$ is necessary to fit for the SN bump in the r-band LC, and we show the r-band LC with and without the inclusion of this stretching factor in the plot. These findings all show that the magnetar model is not viable to describe the LC of GRB 230812B/SN 2023pel, and that an additional source of power is necessary to describe the observed flux. }
    \label{magnetarmodel}
\end{figure}

Given this formulation, we begin by fitting Eq. \ref{equ:mag_combined} to the \textit{Swift} XRT LC, where we convert the 0.3--10 keV flux LC to a rest-frame 0.3–10 keV X-ray luminosity LC through the same method described in Section 3 of \citet{Cano2016}, fitting the power-law decay index to $\alpha = \Gamma_\gamma + 1 = 3.16$ \citep{photonindexGCN}. The best-fit LC is shown in black in Figure \ref{magnetarmodel}, and we derive $L_0 = 2.62^{+0.45}_{-0.28} \times 10^{45} \, \rm{erg \, s^{-1}}$ and $t_0 = 3.75^{+0.51}_{-0.61} \times 10^{4} \, \rm{s}$. This in turn corresponds to $B_0 = 3.37^{+0.91}_{-0.63} \times 10^{15} \, \rm{G}$ and $P_0 = 14.44^{+2.29}_{-1.93} $ ms. We then convert our $r$ band LC into luminosity space, and fit Eq. \ref{equ:mag_combined_norm_SN} to the LC, where the best-fit LC is shown in red in Figure \ref{magnetarmodel}. We derive $L_0 = 1.03\pm0.09 \times 10^{45} \, \rm{erg \, s^{-1}}$, $t_0 = 1.89^{+0.15}_{-0.13} \times 10^{4} \, \rm{s}$, and $\Phi = 8.84^{+0.62}_{-0.63}$, which shows clearly that an additional source of energy is needed to power the SN. These parameters correspond to $B_0 = 1.07^{+0.13}_{-0.12} \times 10^{16} \, \rm{G}$ and $P_0 = 32.41^{+2.67}_{-2.48} $ ms. 

There is therefore a clear discrepancy found in the values derived for the magnetar through independently fitting the X-ray and optical LCs. This shows that the magnetar model under the assumption of a dipolar, unevolving magnetic field, is not viable to satisfactorily describe the observed phases of GRB 230812B/SN 2023pel.

\begin{figure*}
    \centering
    \includegraphics[width = \linewidth]{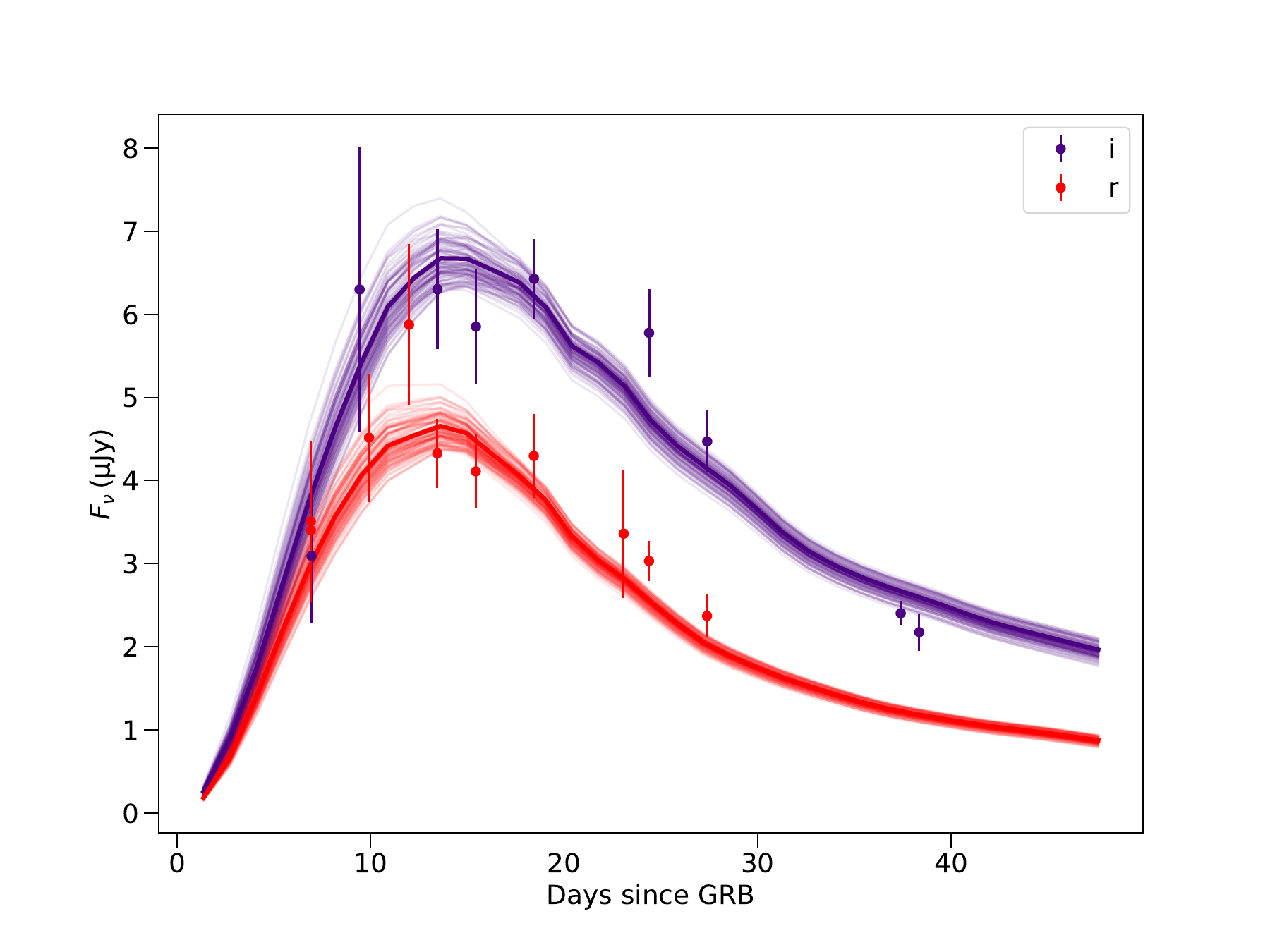}
     \caption{The afterglow and host-subtracted SN flux LC of SN 2023pel, corrected for Galactic extinction. In addition, we plot the best-fit $r$ and $i$-band flux LCs in the observer frame obtained from the best-fit bolometric LC constructed from the A82 radioactive heating SN model, assuming the color and spectral evolution of SN 2023pel is identical to that of SN 2002ap. We derive the best-fit values $M_{\rm{Ni}} = 0.38 \pm 0.01\, \rm{M_{\odot}}$, $\tau_{\rm{m}} = 7.64^{+0.34}_{-0.33}$ days. 100 random a posteriori possible models from the MCMC fitting samples are also plotted with the best fits shown in bold.}
    \label{Nifits}
\end{figure*}

\subsection{Arnett Model and SN parameters}
\label{Arnett}
We then consider the conventional \citet{Arnett1982} model. This model, also known as the radioactive heating model, assumes that the optical emission is due to the decay of $^{56}\rm{Ni}$ to \Co~ and to \Fe~. We use Equation 36 from \citet{Arnett1982} to model the bolometric optical luminosity, assuming full gamma-ray trapping of the ejecta, in addition to further radioactive inputs \citep{Valenti2008}. The $^{56}\rm{Ni}$ mass is a key parameter that can be used to provide insights into both the explosion and the progenitor, and is one of the major SN parameters we fit for in our modeling.

In order to fit the \citet{Arnett1982} model, we begin by isolating the SN flux, by subtracting the best-fit power-law from \S \ref{aganalysis} from the observed photometry after $\rm{T_0} + 5$ days, when the afterglow has faded enough for the SN flux to make relevant contributions to the LC. SN 2023pel's flux LC, corrected for Galactic extinction, is shown in Figure \ref{Nifits}. We then generate semi-analytic bolometric luminosity LC models from \citet{Arnett1982} using the Hybrid Analytic Flux FittEr for Transients (HAFFET; \citealt{Yang+2023}), where the two free parameters are the Nickel mass (${M}_{\rm Ni}$) and the photon diffusion timescale ($\tau_{\rm{m}}$).  $\tau_{\rm{m}}$ is an important parameter that relates to the mass of the total ejecta and the kinetic energy of the explosion, which we will show in \S \ref{spectra}.

Given a bolometric luminosity LC from the models, we need to extract associated $r$ and $i$-band LCs corresponding to the models to compare to our observed photometry. We do so through a similar method as in \citet{Srinivasaragavan2023}, where we use bolometric correction (BC) coefficients at every epoch to convert from a bolometric luminosity LC to individual filter LCs, assuming that the color and spectral evolution of the SN is identical to that of the Type Ic-BL SN 2002ap (see \S \ref{spectra}; \citealt{Mazzali2002}). The BC corrections are defined as:

\begin{equation}
  \text{BC}_{x} = M_{\rm{bol}} - M_{x},
\label{eq:bc}
\end{equation}
where $x$ is the relevant filter, $M_{\rm{bol}}$ is the optical absolute bolometric magnitude, and $M_{x}$ is the absolute magnitude in the relevant filter. For stripped-envelope SNe in the photospheric phase, the $g$-band coefficient \citep{Lyman2014} is:

\begin{equation}
  \text{BC}_{g} = 0.054 - 0.195 \times (g-r) - 0.719 \times (g-r)^{2} .
  \label{eq:bc_se_sl}
\end{equation}

We derive the $g$-band coefficients at every epoch from a SN 2002ap-like LC at $z=0.36$ generated through {\sc SNCosmo}, as well as the $g-r$ and $g-i$ colors. Then, at every epoch, we convert the bolometric luminosity computed from the \citet{Arnett1982} model to a bolometric absolute magnitude, and use Eq. \ref{eq:bc} to compute a $g$-band LC. Finally, we use the associated $g-r$ and $g-i$ colors at every epoch to generate $r$ and $i$-band LCs. After applying the correct distance modulus and K-corrections to these LCs, we compare them to our observed SN flux using Markov Chain Monte Carlo (MCMC) techniques with the \texttt{Python} package {\sc emcee} \citep{emcee} to fit for the nickel mass and the photon diffusion timescale, which have uniform priors corresponding to values derived in literature \citep{Taddia2018, Corsi2016, Corsi2022}. We note that this method (though instead using a SN 1998bw-like LC) was also used to estimate the nickel mass of SN 2022xiw, the SN associated with GRB 221009A, in \citet{Srinivasaragavan2023}. They derived results consistent with those of \citet{Blanchard2023}, who analyzed a \textit{JWST} NIRSpec spectrum to constrain the nickel mass of SN 2022xiw.

The best-fit LCs from our fitting are shown in Figure \ref{Nifits}. We derive ${M}_{\rm Ni} = 0.38 \pm 0.01 \, \rm{M_\odot}$ and $\tau_{\rm{m}} = 7.64^{+0.34}_{-0.33}$. We note that this error is the statistical uncertainty, and there are likely systematic uncertainties that arise from the \citet{Arnett1982} model. Specifically, the assumptions of spherical symmetry along with full gamma-ray trapping of the ejecta play the biggest role in these uncertainties. The $\rm{M}_{\rm Ni}$ we find is consistent with studies of SN 1998bw ($M_{\rm{Ni}} = 0.3$--$0.9 \, \rm{M_{\odot}}$; \citealt{Sollerman2000}). This is expected as the brightness of the SN is about the same as SN 1998bw (see \S \ref{SNbump}), and the nickel mass is a proxy for the brightness of the SN. These values correspond to a peak bolometric luminosity of $L_{\rm{bol}} \sim 1.3 \times 10^{43}\,  \rm{erg \, s^{-1}}$, which is consistent with the average found for the overall GRB-SN sample of $L_{\rm{bol}} = 1 \times 10^{43}\,  \rm{erg \, s^{-1}}$ with dispersion $\sigma = 0.4 \times 10^{43}\,  \rm{erg \, s^{-1}}$ \citep{cano2017}.

We then estimate the photospheric expansion velocity ($v_{\mathrm{ph}}$), through measuring the absorption velocity of the Fe II feature at 5169 $\AA$ from the SN spectrum (shown in Figure \ref{SNSpectrum}), which has been shown to be a good proxy for $v_{\rm{ph}}$ \citep{Modjaz2016}. We use the same method as \citet{Anand2022}, using \sw{SESNSpectraPCA} \citep{SESNSpectraPCA} to smooth the spectrum. We then use \sw{SESNSpectraLib} \citep{Modjaz2016, Liu2016} to fit for the Fe II absorption velocity, by convolving the spectrum with Type Ic SN templates. From our fitting procedure, we estimate that $v_{\rm{ph}} = 11,300 \pm 1,600 \, \rm{km \, s^{-1}}$. This is lower than the velocities expected at peak for Type Ic-BL SNe, but consistent with values derived for spectra taken around the same phase after peak for other Type Ic-BL events \citep{Taddia2018, Modjaz2016}. For a sanity check, we estimate the photospheric velocity for SN 2002ap \citep{Mazzali2002} and SN 1998bw \citep{patat2001} at the phases of the best-fit SNID templates. We find that $v_{\rm{ph}} = 11,900 \pm 1,150 \, \rm{km \, s^{-1}}$ for SN 2002ap, and $v_{\rm{ph}} = 14,100 \pm 800 \, \rm{km \, s^{-1}}$ for SN 1998bw. Therefore, the photospheric velocities we estimate for SN 2023pel are consistent with SN 2002ap at the 1$\sigma$ level at a similar phase, and consistent with SN 1998bw at the 2$\sigma$ level at a later phase corresponding to a best-fit template from SNID.

Given $\tau_m$ and $v_{\rm{ph}}$, we can derive the total mass ejected in SN 2023pel ($M_{\rm{ej}}$), and the total kinetic energy of the explosion ($E_{\rm{KE}}$), using the equations from \citet{Lyman2016}. Assuming that the explosion is a constant density sphere undergoing homologous expansion, $M_{\rm{ej}}$ is described as

\begin{equation}
 M_\mathrm{ej} = \frac{\tau_\mathrm{m}^2\beta c v_\mathrm{sc}}{ 2\kappa_\mathrm{opt}}\textrm{\! ,}
 \label{eq4}
\end{equation}
and $E_{\rm{KE}}$ is described as 
\begin{equation}
  E_\mathrm{KE}  = \frac{3 v_\mathrm{sc}^2 M_\mathrm{ej}}{10}\textrm{\! ,}
 \label{eq:vsc}
\end{equation}
 where $\beta = 13.8$ is a constant, $c$ is the speed of light, $\kappa_{\mathrm{opt}}$ is a constant, average optical opacity, and $v_{\rm{sc}}$ is observationally set to the photospheric velocity $v_{\rm{ph}}$ at maximum light. We note that $\kappa_{\mathrm{opt}}$ for stripped-envelope SNe varies in literature (\citealt{kopt} quotes $\kappa_{\mathrm{opt}} = 0.18$\,cm$^{-2}$\,g$^{-1}$ for Type Ib SNe, and $\kappa_{\mathrm{opt}} = 0.10$\,cm$^{-2}$\,g$^{-1}$ for Type Ic SNe), but we adopt the value used by \citet{Chugai2000}, \citet{Tartaglia2021}, and \citet{Barbarino2020} for stripped-envelope SNe, $\kappa_{\mathrm{opt}} = 0.7$\,cm$^{-2}$\,g$^{-1}$, shown to accurately model observed stripped-envelope SNe in hydrodynamical LCs \citep{Taddia2018a}. Because our observed spectrum is taken 15.5 days after the peak, we cannot use the photospheric velocity we derived earlier to estimate these parameters, and can only derive lower limits. We find $M_{\rm{ej}} > 0.58 \, \rm{M_\odot}$ and $E_{\rm{KE}} > 3.2\times10^{50} \, \rm{erg}$.

\begin{figure*}
    \centering
    \includegraphics[width = 0.49\linewidth]{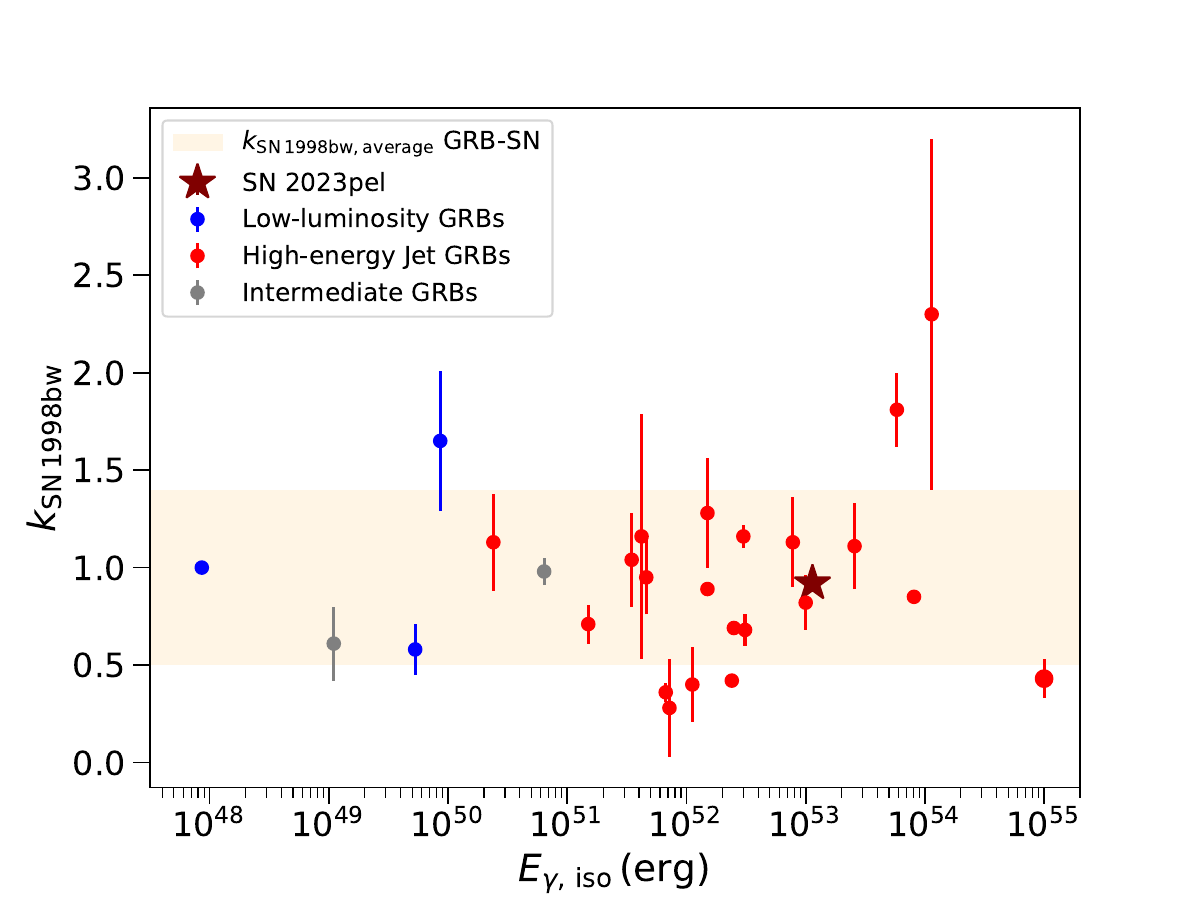}
     \includegraphics[width = 0.49\linewidth]{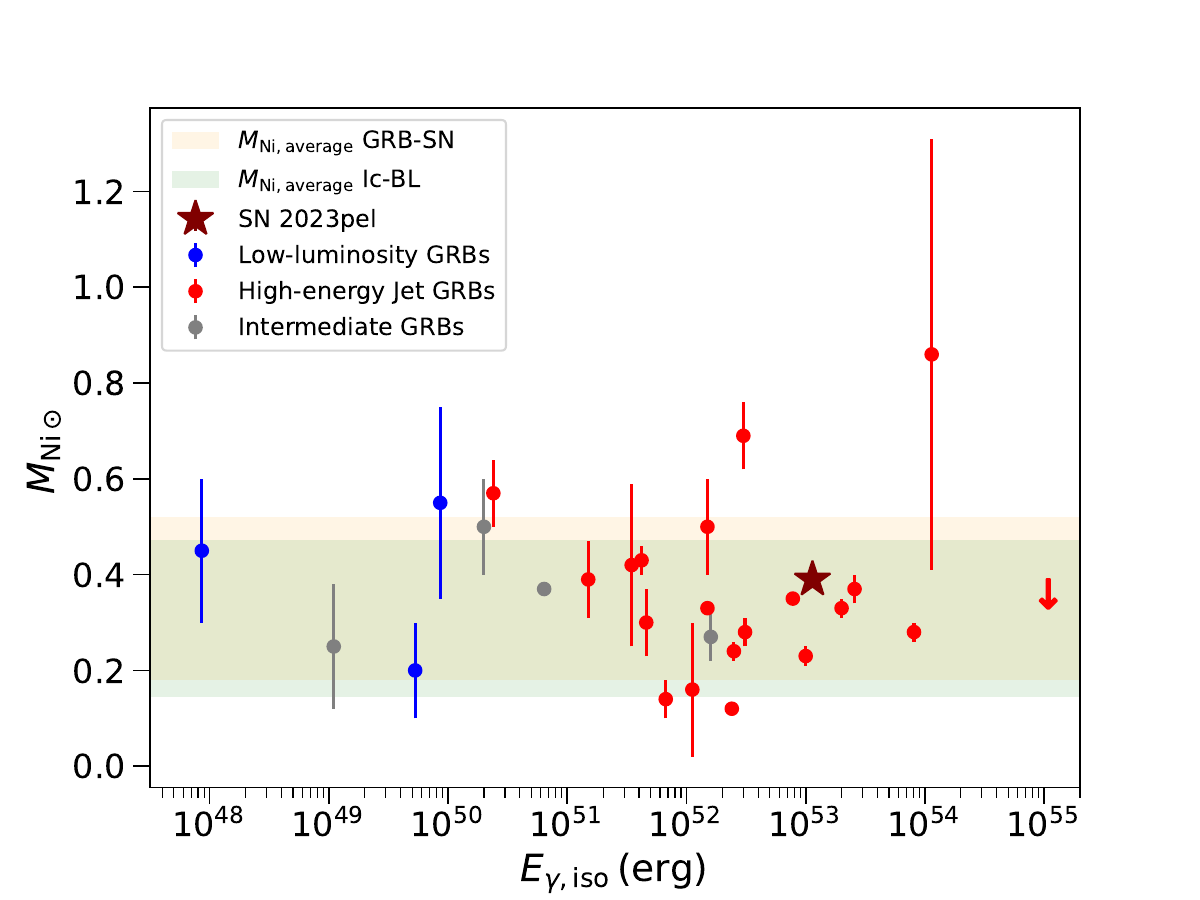}
    \caption{\textit{Left panel}: 
    Modification of Figure 5 from \citet{Srinivasaragavan2023}, where $k_{\rm{SN1998bw}}$ is plotted against the isotropic equivalent gamma-ray energy ($E_{\gamma, \rm{iso}}$) for observationally confirmed GRB-SNe. The plot distinguishes between low-luminosity GRBs, high-energy jet GRBs, and GRBs in the intermediate regime, and also shows the average $k_{\rm{SN1998bw}}$ for GRB-SNe in the plot and its dispersion, without accounting for GRB 230812B/SN 2023pel.  We indicate the results from this work, GRB 230812B/SN 2023pel, with a maroon star. \textit{Right panel}: A similar update of Figure 5 from \citet{Srinivasaragavan2023}, where the $M_{\rm{Ni}}$ in $\rm{M_{\odot}}$ of GRB-SNe is plotted against the isotropic equivalent gamma-ray energy  ($E_{\gamma, \rm{iso}}$) from the prompt emission for the GRBs. The same distinctions for GRBs are made as for the left panel, and we show the results from this work with a maroon star. We also show the average $M_{\rm{Ni}}$, along with their dispersion, for GRB-SN in the plot, with the exception of GRB 230812B/SN 2023pel as well as GRB 221009A/SN 2022xiw, because only a robust upper limit was derived for the event from \citet{Srinivasaragavan2023} We also show the average $M_{\rm{Ni}}$, along with their dispersion for the Type Ic-BL SNe sample not associated with GRBs from \citet{Taddia2018}.}
    \label{GRBSNfigure}
\end{figure*}

In order to derive $v_{\rm{sc}}$ and these parameters, we use the photospheric velocity evolution of the Type Ic-BL SNe sample from  \citet{Modjaz2016} to estimate SN 2023pel's photospheric velocity at maximum light. Using the values from Table 3 in \citet{Modjaz2016}, we find the median photospheric velocities and standard deviations at every time epoch between -15 and 20 days from maximum light, and shift the median and 1$\sigma$ time evolution curves to our derived velocity, $v_{\rm{ph}} = 11,300 \pm 1,600 \, \rm{km \, s^{-1}}$, at 11.3 rest-frame days after SN 2023pel's observed peak. We then extrapolate to the time of maximum light (0 days), and derive $v_{\rm{sc}} = 14,800\pm 7,500 \,\rm{km \, s^{-1}}$, which is broadly consistent with the results from \citet{Hussennot2023} who derive a photospheric velocity at peak of  $v_{\rm{sc}} = 17114\pm 2993 \,\rm{km \, s^{-1}}$. Then, using the derived photospheric velocity at peak, we find $M_{\rm{ej}} = 1.0 \pm 0.6 \, \rm{M_\odot}$ and $E_{\rm{KE}} = 1.3 ^{+3.3}_{-1.2}\times10^{51} \, \rm{erg}$.

\subsection{Comparison to GRB-SN population}
\label{GRBSN}

Here, we contextualize GRB 230812B/SN 2023pel with respect to the overall GRB-SN population. The average values and dispersion of  $k_{\rm{SN1998bw}}$ and $M_{\rm{Ni}}$ for the GRB-SN population are $k_{\rm{SN1998bw}} = 0.95$ with $\sigma = 0.45$, and $M_{\rm{Ni}} = 0.37 \, \rm{M}_{\odot}$  with $\sigma = 0.20 \,\rm{M}_{\odot}$ \citep{cano2017}. Therefore, the $k_{\rm{SN1998bw}}$ and $M_{\rm{Ni}}$ we derive for SN 2023pel are consistent with the overall GRB-SN population. The average values and dispersion of $M_\mathrm{ej}$ and $E_{\rm{KE}}$ for previous GRB-SN are  $M_\mathrm{ej} = 6 \,  \rm{M_\odot}$ with a dispersion $\sigma =  4 \,\rm{M_\odot}$ and $E_{\rm{KE}} = 2.5 \times 10^{52} \, \rm{erg}$, with a dispersion $\sigma = 1.8 \times 10^{52} \, \rm{erg}$ \citep{cano2017}. We find that our derived ejecta masses and kinetic energies are slightly lower than those of the overall population. This is due to the quick time evolution of the SN, leading to a relatively low photon diffusion timescale.


Because most luminous LGRBs are cosmological in origin, their associated SNe are too faint for current optical facilities to detect. However, GRB 221009A provided a rare example of an extremely energetic ($E_{\gamma, \rm{iso}} = 1  \times 10^{55}  \rm{\, erg}$; \citealt{Burns2023}) and luminous ($L_{\gamma, \rm{iso}} = 2.1  \times 10^{54}  \rm{\, erg \, s^{-1}}$;  \citealt{Frederiks2023}) LGRB that was close enough ($z = 0.151$) to study its associated SN. Despite GRB 221009A's large gamma-ray energy, its associated SN's properties were consistent with the overall GRB-SN population \citep{Blanchard2023, Srinivasaragavan2023, Fulton2023, Shrestha+2023, Kann+2023, Levan2023}, and its brightness may have in fact been a little lower than the average seen in the population. The prompt emission from GRB 230812B is not nearly as energetic or luminous as GRB 221009A, but its properties are still on the high end with respect to the GRB-SN population. Therefore, it provides us with another opportunity to understand where SN parameters lie in the higher-energy regime of the GRB-SN population.

In Figure \ref{GRBSNfigure}, we add GRB 230812B/SN 2023pel to a modification of Figure 5 from \citet{Srinivasaragavan2023}, and compare GRBs' $E_{\gamma, \rm{iso}}$ to their associated SNe's $k_{\rm{SN\, 1998bw}}$ and $M_{\rm{Ni}}$, classifying the GRBs into low-luminosity ($L_{\gamma, \rm{iso}} < 10^{48.5} \, \rm{erg\, s^{-1}}$) and high-energy jet GRBs ($L_{\gamma, \rm{iso}} > 10^{49.5} \, \rm{erg\, s^{-1}}$), with events in between labeled as intermediate GRBs. We shift our focus from the average isotropic gamma-ray luminosities for GRBs analyzed in \citet{Srinivasaragavan2023} to their isotropic equivalent gamma-ray energies instead, as it is a  more relevant property for comparisons between the associated SN's parameters connected to its luminosity ($k_{\rm{SN\, 1998bw}}$ and $M_{\rm{Ni}}$). We also plot the average $k_{\rm{SN1998bw}}$ for the GRB-SN population, and $M_{\rm{Ni}}$ for the GRB-SN and overall Type Ic-BL SNe population \citep{Taddia2018} in the Figure. We find that SN 2023pel's properties are very ordinary with respect to the rest of the GRB-SN population.

\citet{Srinivasaragavan2023} also tested for statistical correlations between $L_{\gamma, \rm{iso}}$ and $k_{\rm{SN\, 1998bw}}$, and $L_{\gamma, \rm{iso}}$ and $M_{\rm{Ni}}$, and they found no significant correlations present. We do the same with  $E_{\gamma, \rm{iso}}$, for the high-energy jet GRB population and the entire population, using the Pearson Correlation Coefficient Test. For the nickel mass, we find a coefficient of 0.40 and a p-value of 0.18 for the high-energy jet GRBs, and a coefficient of 0.35 with a p-value of 0.19 for the entire population. For $k_{\rm{SN\, 1998bw}}$, we find a coefficient of -0.16 and a p-value of 0.48 for the high-energy jet GRBs, and a coefficient of -0.15 with a p-value of 0.46 for the entire population. Therefore, there are no significant correlations present again. This, along with other events in literature (see, e.g.,  \citealt{Tanvir2010, Michalowski2018}) suggests a decoupling between the central engine activity that powers relativistic ejecta in GRBs, and SN emission in GRB-SN systems. 

\section{Conclusion}
\label{Conclusion}
We analyze the optical counterpart of GRB 230812B, and determine it possesses a late-time flattening consistent with an associated SN, SN 2023pel. SN 2023pel has a peak $r$-band magnitude of $M_r = -19.46 \pm 0.18$ mag, and has a similar brightness to SN 1998bw, while evolving on quicker timescales. We confirm SN 2023pel is a Type Ic-BL SN through analyzing a spectrum taken by DEIMOS 15.5 days after the SN peak in $r$ band, and confirming broad Fe II and Si II features. We then rule out a millisecond magnetar central engine powering the GRB-SN \citep{Cano2016} in the context of a dipolar, unevolving magnetic field, through an independent fitting of the X-ray and optical LCs. Using the \citet{Arnett1982} radioactive decay model, we find that SN 2023pel has a nickel mass $M_{\rm{Ni}} = 0.38 \pm 0.01 \, \rm{M_\odot}$, which is consistent with both the GRB-SN population and the overall Type Ic-BL SNe population. We derive a photospheric expansion velocity of $v_{\rm{ph}} = 11,300 \pm 1,600 \, \rm{km \, s^{-1}}$ at that phase, and extrapolate a velocity at maximum light of $v_{\rm{ph}} = 14,800 \pm 7500  \, \rm{km \, s^{-1}}$. Using this velocitiy, we derive estimates of the ejecta mass and kinetic energy: $M_{\rm{ej}} = 1.0 \pm 0.6 \, \rm{M_\odot}$ and $E_{\rm{KE}} = 1.3 ^{+3.3}_{-1.2}\times10^{51} \, \rm{erg}$.

Our analysis of GRB 230812B/SN 2023pel shows that SN 2023pel is a rather ordinary SN with respect to the overall GRB-SN population. GRB 230812B/SN 2023pel adds more evidence that the central engine and SN powering mechanisms are decoupled in GRB-SN systems. As optical surveys become more sensitive in the future, we will uncover more GRB-SN events that possess a $E_{\rm{\gamma, iso}}$ between GRB 230812B and GRB 221009A, a region of the parameter space that has been sparsely explored, as seen in Figure \ref{GRBSNfigure}. These events have the potential to help understand why this decoupling between the GRB central engine and SN emission occurs. Therefore, we encourage future studies of nearby, energetic GRB-SNe, in order to continue shedding light on the outstanding open questions of the GRB-SN connection.

\section*{Acknowledgements} 
G.P.S. dedicates this paper to Keagan Jaravata, and to the Alvir and Jaravata families. SN 2023pel will always be connected to the memory of Keagan's life. 

 G.P.S. thanks Isiah Holt for useful discussions on MCMC techniques, and Simi Bhullar for her moral support throughout the paper-writing porcess. The material is based upon work supported by NASA under award number 80GSFC21M0002. B. O. acknowledges useful discussions with Noel Klingler regarding UVOT data analysis and the impact of the ongoing attitude control issues. B. O. gratefully acknowledges support from the McWilliams Postdoctoral Fellowship at Carnegie Mellon University.  M.C.M. was supported in part by NASA ADAP grants 80NSSC21K0649 and 80NSSC20K0288. G.C. Anupama thanks the Indian National Science Academy for support under the INSA Senior Scientist programme. M.W.C. acknowledges support from the National Science Foundation with
grant numbers PHY-2010970 and OAC-2117997. SY is supported by the National Natural Science Foundation of China under Grant No. 12303046. 
 
 These results made use of Lowell Observatory’s Lowell Discovery Telescope (LDT),
formerly the Discovery Channel Telescope. Lowell operates the LDT in partnership with
Boston University, Northern Arizona University, the University of Maryland, and the University of Toledo. Partial support of the LDT was provided by Discovery Communications. LMI was built by Lowell Observatory using funds from the National Science Foundation (AST-1005313). The GROWTH India Telescope (GIT) is a 70-cm telescope with a 0.7-degree field of view, set up by the Indian Institute of Astrophysics (IIA) and the Indian Institute of Technology Bombay (IITB) with funding from Indo-US Science and Technology Forum and the Science and Engineering Research Board, Department of Science and Technology (DST), Government of India.  It is located at the Indian Astronomical Observatory (Hanle), operated by IIA an autonomous institute under DST. We acknowledge funding by the IITB alumni batch of 1994, which partially supports operations of the telescope. Telescope technical details are available at \url{https://sites.google.com/view/growthindia/}.
The 2m Himalayan Chandra telescope (HCT) located at the Indian Optical Observatory (IAO) at Hanle. We thank the staff of IAO, Hanle and CREST, Hosakote, that made these observations possible. HCT observations were carried out under the ToO program of proposal number HCT-2023-C2-P15. The facilities at IAO and CREST are operated by the Indian Institute of Astrophysics, Bangalore.
The Liverpool Telescope is operated on the island of La Palma by Liverpool John Moores University in the Spanish Observatorio del Roque de los Muchachos of the Instituto de Astrofisica de Canarias with financial support from the UK Science and Technology Facilities Council.

Some of the data presented herein were obtained at the W.M. Keck Observatory, which is operated as a scientific partnership among the California Institute of Technology, the University of California and the National Aeronau- tics and Space Administration. The Observatory was made possible by the generous financial support of the W.M. Keck Foundation. The authors wish to recognize and acknowledge the very significant cultural role and reverence that the summit of Mauna Kea has always had within the indigenous Hawaiian community. We are most fortunate to have the opportunity to conduct observations from this mountain. SED Machine is based upon work supported by the National Science Foundation under Grant No. 1106171. Based on observations obtained with the Samuel Oschin Telescope 48-inch and the 60-inch Telescope at the Palomar Observatory as part of the Zwicky Transient Facility project. ZTF is
supported by the National Science Foundation under Grant No. AST-2034437 and a collaboration including Caltech, IPAC, the Weizmann Institute of Science, the Oskar Klein Center at Stockholm University, the University of Maryland, Deutsches Elektronen-Synchrotron and Humboldt University, the TANGO Consortium of Taiwan, the University of Wisconsin at Milwaukee, Trinity College Dublin, Lawrence Livermore National Laboratories, IN2P3, University of Warwick, Ruhr University Bochum and Northwestern University. Operations are conducted by COO, IPAC, and UW. The ZTF forced-photometry service was funded under the Heising-Simons Foundation grant \#12540303 (PI: Graham). The Gordon and Betty Moore Foundation, through both the Data-Driven Investigator Program and a dedicated grant,
provided critical funding for SkyPortal.

\facilities{Zwicky Transient Facility (ZTF), Spectral Energy Distribution Machine (SEDM), Liverpool Telescope (LT), Lowell Discovery Telescope (LDT), GROWTH-India Telesceope (GIT), Himalayan Chandra Telescope (HCT), Keck Deep Imagining Multi-Object Spectrograph (DEIMOS)}

\software{{\sc PyMultinest}, {\sc WOMBAT}, {\sc SESNSpectraPCA}, {\sc SESNSpectraLib}, {\sc Pypeit}, {\sc SNCosmo}, {\sc PySpecKit}, {\sc XSPEC v12.12.0}, 
 {\sc emcee}, {\sc SCAMP}, {\sc Swarp}, {\sc ZOGY},  {\sc SourceExtractor}, {\sc Astro-SCRAPPY}, {\sc solve-field}, {\sc HEASoft v6.29c}, {\sc Fpipe}}


\appendix
\begin{longtable}{ccccc}
\hline
\hline
$t_{\rm{obs}} - T_0$ (days)  & Telescope & Filter & AB mag & Uncertainty \\
\hline
       0.359 & ZTF & $r$ & 18.85 & 0.04 \\ 
        0.359 & ZTF & $r$ & 18.85 & 0.04 \\ 
        0.359 & ZTF & $r$ & 18.83 & 0.02 \\ 
        0.377 & ZTF & $r$ & 18.95 & 0.03 \\ 
        0.377 & ZTF & $r$ & 18.95 & 0.03 \\ 
        0.377 & ZTF & $r$ & 18.96 & 0.01 \\ 
        0.393 & ZTF & $g$ & 19.19 & 0.03 \\ 
        0.393 & ZTF & $g$ & 19.19 & 0.03 \\ 
        0.393 & ZTF & $g$ & 19.21 & 0.01 \\ 
        0.411 & ZTF & $g$ & 19.22 & 0.03 \\ 
        0.411 & ZTF & $g$ & 19.22 & 0.03 \\ 
        0.411 & ZTF & $g$ & 19.29 & 0.03 \\ 
        0.446 & ZTF & $r$ & 19.16 & 0.04 \\ 
        0.446 & ZTF & $r$ & 19.16 & 0.04 \\ 
        0.446 & ZTF & $r$ & 19.16 & 0.02 \\ 
        0.451 & SEDM & $u$ & 19.84 & 0.2 \\ 
        0.455 & SEDM & $g$ & 19.52 & 0.13 \\ 
        0.455 & SEDM & $g$ & 19.52 & 0.13 \\ 
        0.457 & SEDM & $r$ & 19.28 & 0.04 \\ 
        0.46 & ZTF & $r$ & 19.3 & 0.09 \\ 
        0.46 & SEDM & $i$ & 19.1 & 0.05 \\ 
        0.463 & SEDM & $r$ & 19.28 & 0.16 \\ 
        0.489 & ZTF & $r$ & 19.14 & 0.12 \\ 
        0.489 & ZTF & $r$ & 19.14 & 0.12 \\ 
        0.489 & ZTF & $r$ & 19.15 & 0.06 \\ 
        0.49 & SEDM & $u$ & 19.69 & 0.23 \\ 
        0.494 & SEDM & $g$ & 19.68 & 0.14 \\ 
        0.497 & SEDM & $r$ & 19.29 & 0.04 \\ 
        0.497 & ZTF & $r$ & 19.24 & 0.16 \\ 
        0.497 & ZTF & $r$ & 19.24 & 0.16 \\ 
        0.497 & ZTF & $r$ & 19.27 & 0.07 \\ 
        0.5 & SEDM & $i$ & 19.15 & 0.05 \\ 
        0.518 & ZTF & $r$ & 19.28 & 0.1 \\ 
        0.518 & ZTF & $r$ & 19.28 & 0.1 \\ 
        0.518 & ZTF & $r$ & 19.26 & 0.07 \\ 
        0.53 & ZTF & $r$ & 19.26 & 0.12 \\ 
        0.53 & ZTF & $r$ & 19.21 & 0.06 \\ 
        0.531 & SEDM & $g$ & 19.89 & 0.13 \\ 
        0.534 & SEDM & $r$ & 19.4 & 0.03 \\ 
        0.536 & SEDM & $i$ & 19.32 & 0.04 \\ 
        0.832 & GIT & $r$ & 19.98 & 0.05 \\ 
        0.85 & GIT & $g$ & 20.34 & 0.06 \\ 
        0.868 & GIT & $i$ & 19.71 & 0.06 \\ 
        1.368 & SEDM & $g$ & 21.09 & 0.15 \\ 
        1.372 & SEDM & $r$ & 20.76 & 0.07 \\ 
        1.376 & SEDM & $i$ & 20.79 & 0.12 \\ 
        1.86 & GIT & $r$ & 21.38 & 0.05 \\ 
        2.125 & LT & $g$ & 21.77 & 0.24 \\ 
        2.127 & LT & $r$ & 21.16 & 0.14 \\ 
        2.129 & LT & $i$ & 21.15 & 0.18 \\ 
        3.009 & HCT & $r$ & 21.77 & 0.16 \\ 
        4.1 & LT & $r$ & 21.96 & 0.19 \\ 
        4.103 & LT & $i$ & 22.47 & 0.3 \\ 
        6.907 & HCT & $r$ & 22.04 & 0.18 \\ 
        6.91 & GIT & $r$ & 22.06 & 0.1 \\ 
        6.945 & HCT & $i$ & 22.04 & 0.15 \\ 
        9.427 & SEDM & $i$ & 21.67 & 0.23 \\ 
        9.916 & HCT & $r$ & 22.02 & 0.14 \\ 
        11.979 & HCT & $r$ & 21.84 & 0.15 \\ 
        13.445 & LDT & $r$ & 22.14 & 0.08 \\ 
        13.45 & LDT & $i$ & 21.76 & 0.11 \\ 
        15.436 & LDT & $r$ & 22.22 & 0.1 \\ 
        15.439 & LDT & $i$ & 21.86 & 0.11 \\ 
        18.43 & LDT & $r$ & 22.22 & 0.11 \\ 
        18.438 & LDT & $i$ & 21.8 & 0.07 \\ 
        23.074 & LT & $r$ & 22.49 & 0.22 \\ 
        24.382 & LDT & $r$ & 22.6 & 0.07 \\ 
        24.388 & LDT & $i$ & 21.94 & 0.09 \\ 
        27.384 & LDT & $r$ & 22.85 & 0.1 \\ 
        27.394 & LDT & $i$ & 22.21 & 0.08 \\
        37.385 & LDT & $r$ & $> 24.29$ & - \\
        37.390 & LDT & $i$ & 22.86 & 0.06 \\ 
        38.340 & LDT & $r$ & $> 24.12$ & - \\
        38.351 & LDT & $i$ & 22.96 & 0.1 \\  
       
        \hline
\caption{Optical photometry and 1$\sigma$ errors of GRB 230812B/SN 2023pel. The photometry includes contributions from the afterglow and  associated SN, and are all image-subtracted to correct for the host galaxy contribution. All times are in the observer frame, and the magnitudes are not corrected for Galactic extinction.}
 \label{Table1}
\end{longtable}

\bibliography{main}{}
\bibliographystyle{aasjournal}



\end{document}